\documentclass[12pt,prb]{revtex4}

\usepackage{setspace}
\usepackage{graphicx}
\usepackage{dcolumn}
\usepackage{bm}

\begin{document}

\preprint{}

\title{Density matrix theory of transport and gain in quantum cascade lasers in a magnetic field}

\author{Ivana~Savi\'c}
\email{eenis@leeds.ac.uk}
\author{Nenad~Vukmirovi\'c}
\author{Zoran~Ikoni\'c}
 \author{Dragan~Indjin}
 \author{Robert~W.~Kelsall}
 \author{Paul~Harrison}
\affiliation{
School of Electronic and Electrical Engineering,
University of Leeds, Leeds LS2 9JT, United Kingdom
}

\author{Vitomir~Milanovi\'c}
\affiliation{
Faculty of Electrical Engineering, University of Belgrade,
11120 Belgrade, Serbia
}%

\singlespacing

\date{\today}

\begin{abstract}

A density matrix theory of electron transport and optical gain in quantum cascade lasers in an external magnetic field is formulated. Starting from the general quantum kinetic treatment, we describe the intra- and inter-period electron dynamics at the non-Markovian, Markovian and Boltzmann approximation levels. Interactions of electrons with longitudinal optical phonons and classical light field are included in the present description. The non-Markovian calculation for a prototype structure reveals significantly different gain spectra in terms of linewidth and additional polaronic features in comparison to the Markovian and Boltzmann ones. Despite strongly opposed interpretations of the origin of the transport processes in the non-Markovian or Markovian and the Boltzmann approaches, they yield comparable values of the current densities. 

\end{abstract}

\pacs{Valid PACS appear here}
\maketitle

\section{Introduction}

The rapid experimental progress in the field of quantum cascade lasers~\cite{faist} (QCL's) has initiated considerable theoretical activity to explain the underlying physical phenomena and improve their performance by design optimization. To date, both semiclassical and quantum-mechanical theories of carrier transport in QCL's without magnetic field have been proposed.
Semiclassical ones are based on the assumption that coherent processes in QCLs are negligible and electron transport occurs via scattering processes only. They rely on the Boltzmann transport equation, for solution of which a few approaches may be employed. The Monte Carlo method is a stochastic approach which simulates the trajectories of a representative ensemble of carriers.~\cite{iottimc,hansmc,zoranmc} An assumption that the carrier distribution in any particular subband can be approximated by Fermi-Dirac statistics allows the Boltzmann equations to be replaced by simpler and less computationally demanding rate equations.~\cite{zoranre,dragan} The quantum-mechanical models enable the description of phase coherence as well as incoherent scattering processes, and they have been formulated using the density-matrix or nonequilibrium Green's function approach.~\cite{iotti,iotti2,iotti3,greenqcl0,greenqcl,greenqcl2,coherent,inesqcl}
 Comparison of the results obtained with the Boltzmann and density matrix approaches in a mid-infrared QCL
 performed by Iotti at al.~\cite{iotti} showed that quantum corrections to the current density are negligible. However, the analysis of gain spectra in the nonequilibrium Green's function description demonstrated that the negligence of coherences between QCL states results in significantly broader linewidths.~\cite{greenqcl2} Also, it has been argued that coherences are not irrelevant for transport in terahertz (THz) QCL's where the small anticrossing energies allow for resonant tunneling.~\cite{hansdm} Moreover, a very recent study~\cite{coherent} gave an interpretation where the current across QCL's is entirely coherent. In addition to the studies concentrated on QCL's, there is a mounting theoretical evidence for the presence of quantum coherence features in linear absorption spectra and nonlinear ultrafast optical response for intersubband transitions in unbiased quantum wells (QW's).~\cite{ines,ines2,stephan,stephan2}

Furthermore, experimental interest in the QCL performance in a magnetic field stimulated theoretical efforts to describe the influence of magnetic field on the physical processes involved. However, since this research topic emerged recently, very few theoretical studies of QCL's in a magnetic field, compared to the amount of those for QCL's without magnetic field, have been reported. Most of them were focused on the modeling of various scattering rates (electron-longitudinal optical phonon,~\cite{becker,smirnov1,smirnov2,lobecker} electron-electron,~\cite{kempa,kempa2} and interface roughness~\cite{newmagnqcl}) and the calculation of these scattering rates between the upper and the lower laser levels. Modeling of the active region of QCL's, including electron-longitudinal optical (LO) phonon and electron-longitudinal acoustic (LA) phonon scattering, and assuming a unity injection approximation, has also been reported.~\cite{cuba,jelena} Finally, a semiclassical model of the electron transport in a magnetic field based on the Boltzmann equation has been proposed.~\cite{prb} Apart from the work done on QCL's in a magnetic field, a few theoretical investigations of QW systems subjected to a magnetic field, based on both the density matrix and nonequilibrium Green's function approaches, have been reported, and confirmed the importance of quantum coherence effects on the ultrafast time scales.~\cite{magnqw,magnqw2}   

Currently, no experimental or theoretical data on coherent phenomena in QCL's in a magnetic field are available. Since the energy spectra in such structures is discrete, it is reasonable to expect that coherent effects are more significant than for QCL's without magnetic field. 
 The aim of this work is to present a quantum-mechanical theory of transport and gain properties of QCL's in an external magnetic field. For that purpose, we derived quantum kinetics equations for QC structures in a magnetic field, based on the density matrix formalism, which include interaction of electrons with LO phonons and optical field. Furthermore, we obtained the corresponding equations in the Markovian approximation, from which the semiclassical Boltzmann transport equations can be recovered. A comprehensive analysis is performed for an example GaAs/Al$_{0.3}$Ga$_{0.7}$As QCL and nonequilibrium steady state results obtained from all three approaches (quantum kinetic, Markovian and Boltzmann) are compared.

\section{Theoretical considerations}
\label{sec2}

\subsection{Quantum kinetic equations}
\label{subsec1}

We consider electrons in the conduction band of a QCL in a magnetic field applied in the direction perpendicular to QW layers ($z$ axis). Such magnetic field splits the in-plane continuum of quantized subbands into Landau levels (LL's), additionally described by Landau and spin indices.~\cite{landau}
 Within the effective mass and envelope function approximations, and neglecting the spin splitting, the energy of the $j_i$th LL originated from the $m_i$th state (subband), in further considerations denoted with a shorthand subscript $i$, $i=\left|m_i,j_i\right>$, reads
\begin{eqnarray}
E_i=E_{\left|m_i,j_i\right>}=\bar{E}_{m_i}+\left(j_i+\frac{1}{2}\right)\frac{\hbar eB}{m^*},
\label{eq1}
\end{eqnarray}
where $\bar{E}_{m_i}$ is the energy of state $m_i$, $\hbar$ is the reduced Planck's constant, $e$ is the electron charge, $B$ is the applied magnetic field, and $m^*$ is the electron effective mass (taken to be equal to $0.067$ in free electron mass units). For the magnetic vector potential ${\bf A}$ given in the Landau gauge (${\bf A}=Bx{\bf e}_y$), the envelope wave function of the $i$th LL takes the form
\begin{eqnarray}
\Psi_{i,k}({\bf r})=u_{j_i}\left(x-f(k)\right)\psi_{m_i}(z)\frac{e^{iky}}
{\sqrt{L_y}},
\label{eq2}
\end{eqnarray}
where $k$ is the wave vector of the electron, $u_{j_i}\left(x-f(k)\right)$ is the wave function of the harmonic oscillator with $f(k)=k/(eB/\hbar)$, $\psi_{m_i}(z)$ is the wave function of the $m_i$th size-quantized state, and $L_y$ is the dimension of the structure along the $y$ axis.

The model Hamiltonian of the system described above reads
\begin{equation}
\hat{H}=\hat{H}_0+\hat{H}_{el}+\hat{H}_{ep}.
\label{eq3}
\end{equation}
The first term represents the Hamiltonian of noninteracting electrons and phonons in applied electric and magnetic fields. The second and the third term describe electron-light and electron-LO phonon interactions, respectively. In this first step towards formulating a density matrix theory of QCL's in a magnetic field, we do not consider other interaction mechanisms of electrons (with LA phonons, ionized impurities, interface defects, and other electrons).  
The Hamiltonian of free electrons and phonons reads
\begin{equation}
\hat{H}_0=\sum_{i,k}E_i\hat{c}_{i,k}^{\dagger}\hat{c}_{i,k}+\sum_{\bf q}E_{\bf q}\hat{b}_{\bf q}^{\dagger}\hat{b}_{\bf q},
\label{eq4}
\end{equation}
where $E_i$ is the energy of the $i$th electron state, $E_{\bf q}$ is the energy of the phonon of a wave vector ${\bf q}$, and $\hat{c}^{\dagger}_{i,k}$ ($\hat{b}^{\dagger}_{\bf q}$) and $\hat{c}_{i,k}$ ($\hat{b}_{\bf q}$) represent creation and annihilation operators of the electron (phonon), respectively. Electron-light interaction in the dipole approximation is given by the following Hamiltonian
\begin{equation}
\hat{H}_{el}=\sum_{i,j,k,k'}e{\bf A}_R{\bf V}_{ij}^{kk'}\hat{c}_{i,k}^{\dagger}\hat{c}_{j,k'},
\label{eq5}
\end{equation}
where ${\bf A}_R$ represents the magnetic vector potential of a monochromatic light wave incident on a QW structure, given in the Coulomb gauge, and the velocity matrix element is found according to
\begin{equation}
{\bf V}_{ij}^{kk'}=\int d{\bf r}\;\Psi_{i,k}^*({\bf r})\hat{\bf v}_0\Psi_{j,k'}({\bf r}).
\label{eq6}
\end{equation}
The velocity operator for the nonilluminated system $\hat{\bf v}_0$ may be represented as 
\begin{eqnarray}
\hat{\bf v}_0=\frac{1}{m^*}\hat{\bf p}+\frac{e{\bf A}}{m^*},
\label{eq7}
\end{eqnarray}
where $\hat{\bf p}$ is the momentum operator. The Hamiltonian describing electron-phonon interaction can be cast in the form~\cite{kuhn,dm} 
\begin{equation}
\hat{H}_{ep}=\sum_{i,j,k,k',{\bf q}}(g_{k,{\bf q},k'}^{ij}\hat{c}_{i,k}^{\dagger}\hat{b}_{\bf q}\hat{c}_{j,k'}+g_{k,{\bf q},k'}^{ij*}\hat{c}_{j,k'}^{\dagger}\hat{b}_{\bf q}^{\dagger}\hat{c}_{i,k}),
\label{eq8}
\end{equation}
with 
\begin{equation}
g_{k,{\bf q},k'}^{ij}=g_{\bf q}\int d{\bf r}\;\Psi_{i,k}^*({\bf r})e^{i{\bf q}\cdot{\bf r}}\Psi_{j,k'}({\bf r}),
\label{eq9}
\end{equation}
where $g_{\bf q}$ represents the coupling factor. For electron-LO phonon interaction, the coupling factor reads
\begin{equation}
g_{\bf q}=-ie\left[\frac{\hbar\omega_{\text{LO}}}{2V}\left(\epsilon_{\infty}^{-1}-\epsilon_s^{-1}\right)\right]^{1/2}\frac{1}{q},
\label{eq10}
\end{equation}
where the energy of each phonon mode is considered to be approximately constant ($\hbar\omega_{\text{LO}}$), $V$ is the volume, and $\epsilon_{\infty}$ and $\epsilon_s$ are high-frequency and static permittivity, respectively.
In the case of QW's in a magnetic field, 
the phonon coupling factor $g_{k,{\bf q},k'}^{ij}$ may be written as
\begin{eqnarray}
g_{k,{\bf q},k'}^{ij}=g_{\bf q}\int d{\bf r}\;u_{j_i}^*\left(x-f(k)\right)\frac{e^{-iky}}{\sqrt{L_y}}\psi_{m_i}^*(z)e^{i(q_xx+q_yy+q_zz)}u_{j_j}\left(x-f(k')\right)\frac{e^{ik'y}}{\sqrt{L_y}}\psi_{m_j}(z)\nonumber\\
=g_{\bf q}\int dy\;\frac{e^{-i(k-q_y-k')y}}{L_y}\int dx\;u_{j_i}^*\left(x-f(k)\right)e^{iq_xx}u_{j_j}\left(x-f(k')\right)\int dz\;\psi_{m_i}^*(z)e^{iq_zz}\psi_{m_j}(z)\nonumber\\=g_{\bf q}\delta_{k',k-q_y}H_{j_ij_j}(k,k',q_x)G_{m_im_j}(q_z),
\label{eq11}
\end{eqnarray}
where $H_{j_ij_j}(k,k',q_x)=\int dx\; u_{j_i}^*\left(x-f(k)\right)e^{iq_xx}u_{j_j}\left(x-f(k')\right)$ is the lateral overlap integral and $G_{m_im_j}(q_z)=\int dz\; \psi_{m_i}^*(z)e^{iq_zz}\psi_{m_j}(z)$ is the form factor.

In the density matrix approach, single particle density matrices like the intraband electron density matrices $f_{i_1i_2,k}=\left<\hat{c}_{i_1,k}^{\dagger}\hat{c}_{i_2,k}\right>$ or the phonon occupation number $n_{\bf q}=\left<\hat{b}_{\bf q}^{\dagger}\hat{b}_{\bf q}\right>$ represent fundamental physical quantities. Their diagonal elements determine the occupation probabilities of the states, while the nondiagonal elements correspond to the electron polarizations between two states, and are related to the property of quantum-mechanical coherence (superposition). In this work, a thermal equilibrium of phonons is assumed, hence equations of motion for electron density matrices are sufficient for the description of the system.

In the derivation of the time evolution of single particle density matrices, one starts with the Heisenberg equation of motion.~\cite{kuhn,dm}
The time evolution due to the Hamiltonian of noninteracting electrons and phonons is given as~\cite{kuhn,dm}
\begin{equation}
\frac{d}{dt}f_{i_1i_2,k}\left|_{\hat{H}_0}\right.=\frac{1}{i\hbar}(E_{i_2}-E_{i_1})f_{i_1i_2,k}
\label{eq12}
\end{equation}
The equation of motion in the case of interaction of electrons with light polarized in the $z$-direction reads~\cite{kuhn,dm,sandra}
\begin{eqnarray}
\frac{d}{dt}f_{i_1i_2,k}\left|_{\hat{H}_{el}}\right.=\frac{e}{i\hbar}\sum_{i_3}A_R\left(V_{i_2i_3}f_{i_1i_3,k}-V_{i_3i_1}f_{i_3i_2,k}\right),
\label{eq13}
\end{eqnarray}
with the $z$-component of the velocity matrix element
\begin{equation}
V_{i_1i_2}=\frac{i}{\hbar}(E_{i_1}-E_{i_2})d_{i_1i_2}\delta_{j_{i_1},j_{i_2}},
\label{eq14}
\end{equation}
where $d_{i_1i_2}$ is the matrix element of the operator of the $z$-coordinate. In the present analysis we do not consider any optical-cavity effect and look for nonequilibrium steady state populations and polarizations when $A_R=0$. However, electron-light interaction is essential for the calculation of optical gain.

In the quantum kinetics equations for electron-phonon interaction, phonon-assisted matrices, given with expectation values of $3$ operators $s_{k,{\bf q},k'}^{i_1i_2}=\left<\hat{c}_{i_1,k}^{\dagger}\hat{b}_{\bf q}\hat{c}_{i_2,k'}\right>$, appear, which correlate an initial state consisting of one electron in the state $i_2,k'$ and a phonon with a wave vector ${\bf q}$ to a final state with only one electron in the state $i_1,k$.~\cite{kuhn,dm}
Furthermore, the temporal evolution for the phonon-assisted matrices involves expectation values of four operators, and so on. The resulting infinite hierarchy of equations needs to be truncated in order to access the problem numerically. 
The first order contribution, obtained by neglecting all correlations between electrons and phonons in the spirit of the correlation expansion approach~\cite{kuhn,dm} ($s_{k,{\bf q},k'}^{i_1i_2}\approx \left<\hat{c}_{i_1,k}^{\dagger}\hat{c}_{i_2,k}\right>\left<\hat{b}_{{\bf q}_{xz}}\right>\delta_{k',k}\delta_{q_y,0}=f_{i_1i_2,k}B_{{\bf q}_{xz}}\delta_{k',k}\delta_{q_y,0}$), vanishes if a thermal equilibrium of phonons is assumed.~\cite{ines} The next order in the hierarchy is obtained by taking into account deviations of the phonon-assisted density matrices from the first order factorization $\delta s_{k,{\bf q},k'}^{i_1i_2}=s_{k,{\bf q},k'}^{i_1i_2}-f_{i_1i_2,k}B_{{\bf q}_{xz}}\delta_{k',k}\delta_{q_y,0}$. Then, the following equations for $\delta s_{k,{\bf q},k'}^{i_1i_2}$ are obtained~\cite{kuhn,dm}
\begin{eqnarray}
\frac{d}{dt}\delta s_{k,{\bf q},k'}^{i_1i_2}=\frac{1}{i\hbar}(E_{i_2}+\hbar\omega_{\text{LO}}-E_{i_1})\delta s_{k,{\bf q},k'}^{i_1i_2}-\gamma_{ph}\delta s_{k,{\bf q},k'}^{i_1i_2}\nonumber\\+\frac{1}{i\hbar}\sum_{i_4,i_5}g_{k,{\bf q},k'}^{i_5i_4*}\left[(n_0+1)f_{i_1i_5,k}(\delta_{i_4,i_2}-f_{i_4i_2,k'})-n_0f_{i_4i_2,k'}(\delta_{i_1,i_5}-f_{i_1i_5,k})\right],\nonumber\\
\frac{d}{dt}f_{i_1i_2,k}\left|_{\hat{H}_{ep}}\right.=\frac{1}{i\hbar}\sum_{i_3,k',{\bf q}}\left(g_{k,{\bf q},k'}^{i_2i_3}\delta s_{k,{\bf q},k'}^{i_1i_3}+g_{k',{\bf q},k}^{i_3i_2*}\delta s_{k',{\bf q},k}^{i_3i_1*}-g_{k',{\bf q},k}^{i_3i_1}\delta s_{k',{\bf q},k}^{i_3i_2}-g_{k,{\bf q},k'}^{i_1i_3*}\delta s_{k,{\bf q},k'}^{i_2i_3*}\right),
\label{eq15}
\end{eqnarray}
where $n_0$ denotes the equilibrium phonon density given by the Bose-Einstein factor. The terms in the equation for $\delta s_{k,{\bf q},k'}^{i_1i_2}$ are due to the Hamiltonian of free electrons, and of electron-LO phonon interaction, respectively. The equations for phonon-assisted matrices should, in principle, contain a term which describes their time evolution due to electron-light interaction, here relevant only for the calculation of linear optical gain. However, the coupling of the light field to the phonon-assisted matrices in QW's is a higher-order effect~\cite{magnqw} and may be neglected.~\cite{stephan,stephan2} Its inclusion for complex structures like QCL's in a magnetic field would result in a computationally inaccessible task.~\cite{magnqw}

Insertion of higher order terms in the equations for phonon-assisted density matrices should be performed in a self-consistent manner,~\cite{haug} however, in the system considered, with several subbands and LL's originating from them in each period of the cascade, this would be extremely computationally involved.~\cite{kuhn,dm} Conversely, discarding these effects leads to numerical instabilities in the actual computation. Therefore, a phenomenological damping constant $\gamma_{ph}$ was introduced, representing higher order correlations.~\cite{stephan,stephan2} We have verified that the convergence of our results may be achieved for sufficiently large values of $\gamma_{ph}$ ($\sim 1$~meV). 

 It is shown in Appendix A that, in the present description, $f_{i_1i_2,k}$ is constant for all values of the wave vector $k$, and may be expressed as $f_{i_1i_2,k}=\alpha_Bn_{i_1i_2}$, where $\alpha_B=\pi\hbar/eB$, and $n_{i_1i_2}=\sum_{k'}f_{i_1i_2,k'}/L_xL_y$. The diagonal element $n_{ii}$ represents the electron sheet density in the $i$th LL. From the derivation given in Appendix A, quantum kinetics equations including all the aforementioned interactions amount to
\begin{eqnarray}
\frac{d}{dt}n_{i_1i_2}=\frac{1}{i\hbar}(E_{i_2}-E_{i_1})n_{i_1i_2}+\frac{e}{i\hbar}\sum_{i_3}A_R\left(V_{i_2i_3}n_{i_1i_3}-V_{i_3i_1}n_{i_3i_2}\right)\nonumber\\+\frac{1}{i\hbar}\sum_{i_3,i_4,i_5}\left(W_{i_2i_3i_4i_5}\delta K_{i_1i_3i_4i_5}+W_{i_3i_2i_5i_4}^*\delta K_{i_3i_1i_5i_4}^*-W_{i_3i_1i_5i_4}\delta K_{i_3i_2i_5i_4}-W_{i_1i_3i_4i_5}^*\delta K_{i_2i_3i_4i_5}^*\right),\nonumber\\
\frac{d}{dt}\delta K_{i_1i_2i_4i_5}=\frac{1}{i\hbar}(E_{i_2}+\hbar\omega_{\text{LO}}-E_{i_1})\delta K_{i_1i_2i_4i_5}-\gamma_{ph}\delta K_{i_1i_2i_4i_5}\nonumber\\+\frac{1}{i\hbar}\left[(n_0+1)n_{i_1i_5}(\delta_{i_4,i_2}-\alpha_Bn_{i_4i_2})-n_0n_{i_4i_2}(\delta_{i_1,i_5}-\alpha_Bn_{i_1i_5})\right],\nonumber\\
W_{i_1i_3i_4i_5}=\frac{e^2\hbar\omega_{\text{LO}}}{8\pi^2}(\epsilon_{\infty}^{-1}-\epsilon_s^{-1})\sum_{i_3i_4}\int_{q_{xy=0}}^{\infty}\int_{q_z=-\infty}^{\infty}q_{xy}dq_{xy}dq_z\;\nonumber\\\frac{1}{q_{xy}^2+q_z^2}|H_{j_{i_1}j_{i_3}}(q_{xy})||H_{j_{i_5}j_{i_4}}(q_{xy})|G_{m_{i_1}m_{i_3}}(q_z)G_{m_{i_5}m_{i_4}}^*(q_z)\delta_{j_{i_1}+j_{i_4},j_{i_3}+j_{i_5}},
\label{eq16}
\end{eqnarray}
with the quantities $\delta K_{i_1i_2i_4i_5}$ associated with the phonon-assisted matrices $\delta s_{k,{\bf q},k-q_y}^{i_1i_2}$ through 
\begin{eqnarray}
\delta s_{k,{\bf q},k-q_y}^{i_1i_2}=\sum_{i_4,i_5}g_{\bf q}^*H_{j_{i_5}j_{i_4}}^*(k,k-q_y,q_x)G_{m_{i_5}m_{i_4}}^*(q_z)\delta K_{i_1i_2i_4i_5}.
\label{eq17}
\end{eqnarray}
The quantum-kinetic dynamics is essentially non-Markovian, since the time evolution of density matrix elements depends on their values at earlier times i.e. on the memory of the system.

From discussion in Appendix A it follows that the time evolution of electron populations and polarizations in the Markovian approximation may take the form
\begin{eqnarray}
\frac{d}{dt}n_{i_1i_2}=\frac{1}{i\hbar}(E_{i_2}-E_{i_1})n_{i_1i_2}+\frac{e}{i\hbar}\sum_{i_3}A_R\left(V_{i_2i_3}n_{i_1i_3}-V_{i_3i_1}n_{i_3i_2}\right)\nonumber\\+\sum_{i_3i_4i_5}\left[-\Gamma_{i_2i_3i_4i_5}^{out}n_{i_1i_5}(\delta_{i_4,i_3}-\alpha_Bn_{i_4i_3})-\Gamma_{i_1i_3i_4i_5}^{out\;*}n_{i_2i_5}^*(\delta_{i_4,i_3}-\alpha_Bn_{i_4i_3}^*)\right.\nonumber\\\left.+\Gamma_{i_2i_3i_4i_5}^{in}n_{i_4i_3}(\delta_{i_1,i_5}-\alpha_Bn_{i_1i_5})+\Gamma_{i_1i_3i_4i_5}^{in\;*}n_{i_4i_3}^*(\delta_{i_2,i_5}-\alpha_Bn_{i_2i_5}^*)\right],\nonumber\\
\Gamma_{i_1i_3i_4i_5}^{out}=\frac{\pi}{\hbar}\left[\delta(-E_{i_5}+\hbar\omega_{\text{LO}}+E_{i_4})W_{i_1i_3i_4i_5}(n_0+1)+\delta(-E_{i_5}-\hbar\omega_{\text{LO}}+E_{i_4})W_{i_3i_1i_5i_4}^*n_0\right],\nonumber\\
\Gamma_{i_1i_3i_4i_5}^{in}=\frac{\pi}{\hbar}\left[\delta(-E_{i_5}+\hbar\omega_{\text{LO}}+E_{i_4})W_{i_1i_3i_4i_5}n_0+\delta(-E_{i_5}-\hbar\omega_{\text{LO}}+E_{i_4})W_{i_3i_1i_5i_4}^*(n_0+1)\right].
\label{eq18}
\end{eqnarray}
Terms $\Gamma_{i_1i_3i_4i_5}^{out/in}$ have similar form as scattering rates in the Boltzmann approach, and hence may be referred to as generalized out/in scattering rates. The Markovian approximation neglects the memory time of a scattering process, which is related to energy-time uncertainty.~\cite{kuhn,dm,haug} Scattering and dephasing processes are then restricted only to energy conserving transitions between single-particle states. For discrete energy spectra in QW's in a magnetic field, the electron-LO phonon interaction is thus almost fully suppressed, if broadening is not taken into account. Therefore, a Lorentzian with the full width at half maximum (FWHM) of $\gamma\sim 1$~meV was used to model the LL broadening in the Markovian description. 
In the semiclassical limit, which may be obtained by neglecting nondiagonal matrix elements,~\cite{kuhn,dm} the derived Markovian equations reduce to the Boltzmann equations given in Ref.~\onlinecite{prb}.

Due to the periodicity of the QCL structure, its energy states are invariant upon translation per potential drop across a period, while the wave functions are invariant upon translation per period length. Therefore, each period has an identical set of $N$ LL's, with identical density matrix elements ($n_{i_1i_2}=n_{(i_1+kN)(i_2+kN)}$, $k=0,\pm1,\pm2,...$), and phonon-assisted matrices ($\delta K_{i_1i_2i_3i_4}=\delta K_{(i_1+kN)(i_2+kN)(i_3+kN)(i_4+kN)}$). This also accounts for the quantities characterizing scattering processes ($\Gamma_{i_1i_2i_3i_4}=\Gamma_{(i_1+kN)(i_2+kN)(i_3+kN)(i_4+kN)}$, $W_{i_1i_2i_3i_4}=W_{(i_1+kN)(i_2+kN)(i_3+kN)(i_4+kN)}$) and the velocity operator ($V_{i_1i_2}=V_{(i_1+kN)(i_2+kN)}$). Since the wave functions are well localized within their periods, the tight-binding description may be introduced, by accounting for the interaction between the nearest neighboring periods only. 
Hence, we consider the density matrix elements which couple LL's within one period, as well as the elements which couple those LL's with LL's belonging to the nearest neighboring periods. Also, we take into account those quantities $W_{i_{1,2}i_3i_4i_5}$ and $\Gamma_{i_{1,2}i_3i_4i_5}$ with the properties $i_{1,2}-i_3=0,\pm1$ and $i_4-i_5=0,\pm1$, see Eq.~(\ref{eq16}), and write Eqs.~(\ref{eq16}) and (\ref{eq18}) for all possible combinations of indices $i_1-i_5$ which satisfy these conditions. After exploiting the property of shift-invariance of all the aforementioned quantities, the system of equations in the quantum-kinetic/Markovian approach may be reduced to contain only the density matrix elements of interest. Again, the Boltzmann expressions may be recovered from the Markovian ones.

In the quantum-kinetic and Markovian representations, the number of the density matrix elements to be calculated is of the order of $N^2$, and the number of the quantities associated with phonon assisted matrices (the quantum kinetics case only) and scattering rates is of the order of $N^4$. Obviously, the calculation of population and polarization dynamics for the QCL's with many energy states and LL's stemming from them is extremely challenging. Therefore, in our analysis, we restrict to the case of a QCL with a small number of energy levels per period, and subjected to relatively large magnetic fields, characterized by a small number of LL's stemming from those levels which are relevant for the transport. Here we took $10$ lowest LL indices, after checking that this number of LL's is sufficient for the considered structure. 

Stationary solution of Eqs.~(\ref{eq16}) and (\ref{eq18}) is found by tracking their time evolution, starting from an initial condition that all electrons are in the fundamental ground state LL (and hence all the polarizations and phonon-assisted matrices are equal to zero), and integrating in time until the steady state is reached. This method proved to be extremely reliable in terms of convergence for solving large systems of nonlinear equations, in contrast to gradient-based methods. The integration is performed by using a Runge-Kutta method with adaptive step size control, which considerably speeds up the process. 

Since the quantities associated to scattering processes $W_{i_1i_2i_3i_4}$ and $\Gamma_{i_1i_2i_3i_4}$ are different from zero only if the condition $j_{i_1}+j_{i_3}=j_{i_2}+j_{i_4}$ is fulfilled, our choice of an initial condition leads to the steady-state solution in which the polarizations $n_{i_1i_2}$ are not equal to zero only if $j_{i_1}=j_{i_2}$. Although this result may, at first, seem to be a peculiarity of the initial condition, it can also be regarded as a solution of the reduced description of the systems of Eqs.~(\ref{eq16}) or (\ref{eq18}), which includes only those density matrix elements $n_{i_1i_2}$ with $j_{i_1}=j_{i_2}$. Careful examination of Eqs.~(\ref{eq16}) and (\ref{eq18}) suggests that, if the terms such as $\alpha_Bn_{i_4i_3}$, $j_{i_4}\ne j_{i_3}$, are much smaller than $1$ (i.e. $\delta_{i_4i_3}$ for $i_4=i_3$), the terms such as $n_{i_1i_5}$, $j_{i_1}\ne j_{i_5}$, do not influence the time evolution of the elements $n_{i_1i_2}$, $j_{i_1}=j_{i_2}$. Therefore, in the lowest order approximation, the analysis of the density matrix dynamics may be limited to the terms $n_{i_1i_2}$, $j_{i_1}=j_{i_2}$.

\subsection{Current density}

The current density may be estimated from the expectation value of the carrier drift velocity $\hat{v}$~\cite{greenqcl}
\begin{equation}
J=\left<\hat{J}\right>=-\frac{e}{V}\left<\hat{v}\right>.
\label{eq19}
\end{equation}
In the density matrix formalism, the drift velocity may be calculated according to~\cite{iotti3}
\begin{equation}
\left<\hat{v}\right>=\sum_{i_1,i_2,k}v_{i_1i_2}f_{i_2i_1,k},
\label{eq20}
\end{equation}
where $v_{i_1i_2}$ is the drift velocity matrix element given with
\begin{equation}
v_{i_1i_2}=\left<i_1\right|\hat{v}\left|i_2\right>=\frac{i}{\hbar}\left<i_1\right|[\hat{H},\hat{z}]\left|i_2\right>.
\label{eq21}
\end{equation}
The drift velocity density matrix elements then reads 
\begin{equation}
v_{i_1i_2}=\frac{i}{\hbar}(E_{i_1}-E_{i_2})d_{i_1i_2}+\frac{1}{m^*}eA_R\delta_{i_1,i_2}.
\label{eq22}
\end{equation}
The first term is due to the Hamiltonian of noninteracting electrons, while the second one is due to the electron-light interaction. Since the scattering potential of any interaction (electron-phonon, electron-electron etc.) depends only on the position $\hat{\bf r}$, but not on the momentum $\hat{\bf p}$, it follows that $[\hat{H}_{ep},\hat{z}]=0$, and its contribution to the drift velocity vanishes.  
Starting from Eqs.~(\ref{eq19}) and (\ref{eq20}), and using the assumption that only the density matrix elements between LL's within a period or LL's localized in that period and its nearest neighbors are not zero, the current density finally may be written in the form
\begin{equation}
J=-\frac{e}{d}\sum_{i_1,i_2=1}^N\left[v_{i_1i_2}n_{i_2i_1}+v_{i_1(i_2+N)}n_{(i_2+N)i_1}+v_{(i_2+N)i_1}n_{i_1(i_2+N)}\right],
\label{eq23}
\end{equation}
where $d$ is the length of a period.
The Boltzmann expression for the current density may be derived from Eq.~(\ref{eq23}) by representing the non-diagonal density matrix elements, in the first order approximation, in terms of the diagonal ones~\cite{jpc}
\begin{eqnarray}
 J=\frac{e}{d}\left\{\sum_{\stackrel{i,f=1}{(i<f)}}^N(z_f-z_i)\left[n_iW_{if}(1-\alpha_Bn_f)-n_fW_{fi}(1-\alpha_Bn_i)\right]\right.\nonumber\\\left.+\sum_{i,f=1}^N(z_{f+N}-z_i)\left[n_iW_{i(f+N)}(1-\alpha_Bn_f)-n_fW_{(f+N)i}(1-\alpha_Bn_i)\right]\right\},
\label{eq24}
\end{eqnarray}
where $n_i=n_{ii}$ represents the population of the $i$th LL and $z_i=d_{ii}=\left<i\right|z\left|i\right>$.

\subsection{Gain spectra}

The gain spectra in the quantum-kinetic (non-Markovian) description and in the Markovian approximation may be estimated from the linear response of nonequilibrium stationary populations and polarizations to a small optical perturbation. The relationship between the linear variations in the polarization due to the applied optical field $\Delta P(\omega)$ and the current density $\Delta J(\omega)$ gives the following expression for the susceptibility~\cite{greenqcl}
\begin{equation}
\chi(\omega)=\frac{\Delta P(\omega)}{\epsilon_0E(\omega)}=-\frac{i}{\epsilon_0}\frac{\Delta J(\omega)}{\omega E(\omega)}.
\label{eq25}
\end{equation}
The gain coefficient then may be found from~\cite{haug} 
\begin{equation}
g(\omega)=-\frac{\omega}{c}\frac{\text{Im}[\chi(\omega)]}{n},
\label{eq26}
\end{equation}
where $n$ is the refractive index of the system material. 

If Fourier transform of the electric field of light is given with
\begin{equation}
{\bf E}(t)={\bf e}_z\int\frac{d\omega}{2\pi}\;E(\omega)e^{-i\omega t},
\label{eq27}
\end{equation}
then Fourier transform of the corresponding magnetic vector potential in the Coulomb gauge is represented as
\begin{eqnarray}
{\bf A}_R(t)={\bf e}_z\int\frac{d\omega}{2\pi}\;\frac{E(\omega)}{i\omega}e^{-i\omega t}.
\label{eq28}
\end{eqnarray}
The linear changes of intra-period elements in the frequency domain $\Delta n_{i_1i_2}(\omega)$, for the quantum kinetics case, may be written as
\begin{eqnarray}
-i\omega\Delta n_{i_1i_2}(\omega)=\frac{E_{i_2}-E_{i_1}}{i\hbar}\Delta n_{i_1i_2}(\omega)+\frac{e}{i\hbar}A_R(\omega)\sum_{i_3}(V_{i_2i_3}n_{i_1i_3}^0-V_{i_3i_1}n_{i_3i_2}^0)\nonumber\\+\frac{1}{i\hbar}\sum_{i_3,i_4,i_5}\left(W_{i_2i_3i_4i_5}\Delta K_{i_1i_3i_4i_5}(\omega)+W_{i_3i_2i_5i_4}^*\Delta K_{i_3i_1i_5i_4}^*(-\omega)\right.\nonumber\\\left.-W_{i_3i_1i_5i_4}\Delta K_{i_3i_2i_5i_4}(\omega)-W_{i_1i_3i_4i_5}^*\Delta K_{i_2i_3i_4i_5}^*(-\omega))\right),\nonumber\\
\Delta K_{i_1i_2i_3i_4}(\omega)=-\frac{1}{E_{i_2}+\hbar\omega_{\text{LO}}-E_{i_1}-\hbar\omega-i\hbar\gamma_{ph}}\nonumber\\\left\{(n_0+1)\left[\Delta n_{i_1i_4}(\omega)\delta_{i_3,i_2}-\alpha_B(n_{i_1i_4}^0\Delta n_{i_3i_2}(\omega)+n_{i_3i_2}^0\Delta n_{i_1i_4}(\omega))\right]\right.\nonumber\\\left.-n_0\left[\Delta n_{i_3i_2}(\omega)\delta_{i_1,i_4}-\alpha_B(n_{i_1i_4}^0\Delta n_{i_3i_2}(\omega)+n_{i_3i_2}^0\Delta n_{i_1i_4}(\omega))\right]\right\},
\label{eq29}
\end{eqnarray}
where $n_{i_1i_2}^0$ represents the steady-state value of the density matrix element between LL's $i_1$ and $i_2$.
Similar equations which include all possible combinations of $i_1-i_5$ as discussed in Subsection~\ref{subsec1} also need to be taken into account. In the Markovian case, taking into consideration the property of the density matrix elements that $n_{i_2i_1}(t)=n_{i_1i_2}^*(t)$, the equations of motion of intra-period density matrix elements transform to the frequency domain according to
\begin{eqnarray}
-i\omega\Delta n_{i_1i_2}(\omega)=\frac{E_{i_2}-E_{i_1}}{i\hbar}\Delta n_{i_1i_2}(\omega)+\frac{e}{i\hbar}A_R(\omega)\sum_{i_3}(V_{i_2i_3}n_{i_1i_3}^0-V_{i_3i_1}n_{i_3i_2}^0)\nonumber\\+\sum_{i_3,i_4,i_5}\left[-\Gamma_{i_2i_3i_4i_5}^{out}(\Delta n_{i_1i_5}(\omega)\delta_{i_4,i_3}-\alpha_B(n_{i_1i_5}^0\Delta n_{i_4i_3}(\omega)+n_{i_4i_3}^0\Delta n_{i_1i_5}(\omega)))\right.\nonumber\\\left.-\Gamma_{i_1i_3i_4i_5}^{out\;*}(\Delta n_{i_5i_2}(\omega)\delta_{i_3,i_4}-\alpha_B(n_{i_5i_2}^0\Delta n_{i_3i_4}(\omega)+n_{i_3i_4}^0\Delta n_{i_5i_2}(\omega)))\right.\nonumber\\\left.+\Gamma_{i_2i_3i_4i_5}^{in}(\Delta n_{i_4i_3}(\omega)\delta_{i_1,i_5}-\alpha_B(n_{i_1i_5}^0\Delta n_{i_4i_3}(\omega)+n_{i_4i_3}^0\Delta n_{i_1i_5}(\omega)))\right.\nonumber\\\left.+\Gamma_{i_1i_3i_4i_5}^{in\;*}(\Delta n_{i_3i_4}(\omega)\delta_{i_5,i_2}-\alpha_B(n_{i_5i_2}^0\Delta n_{i_3i_4}(\omega)+n_{i_3i_4}^0\Delta n_{i_5i_2}(\omega)))\right].
\label{eq30}
\end{eqnarray}
Current density in the frequency domain in both the non-Markovian and Markovian descriptions may be obtained from
\begin{eqnarray}
\Delta J(\omega)=-\frac{e}{d}\left\{\sum_{i_1=1}^N\frac{1}{m^*}eA_R(\omega)n_{i_1i_1}^0+\sum_{i_1,i_2=1}^N\left[\frac{i}{\hbar}(E_{i_1}-E_{i_2})d_{i_1i_2}\Delta n_{i_2i_1}(\omega)\right.\right.\nonumber\\\left.\left.+\frac{i}{\hbar}(E_{i_1}-E_{i2+N})d_{i_1(i_2+N)}\Delta n_{(i_2+N)i_1}(\omega)+\frac{i}{\hbar}(E_{i_2+N}-E_{i_1})d_{i_1(i_2+N)}\Delta n_{i_1(i_2+N)}(\omega)\right]\right\}.
\label{eq31}
\end{eqnarray}

In the Boltzmann description, the gain coefficient may be given as 
\begin{eqnarray}
g(\omega)=\frac{\pi e^2}{n\epsilon_0c\omega\hbar^2d}\sum_{i,f=1}^Nn_i\left[(E_i-E_f)^2\text{sgn}(E_i-E_f)d_{m_i,m_f}^2\delta(|E_i-E_f|-\hbar\omega)\right.\nonumber\\\left.+(E_i-E_{f+N})^2\text{sgn}(E_i-E_{f+N})d_{m_i,m_{f+N}}^2\delta(|E_i-E_{f+N}|-\hbar\omega)\right.\nonumber\\\left.+(E_{i+N}-E_f)^2\text{sgn}(E_{i+N}-E_f)d_{m_{i+N},m_f}^2\delta(|E_{i+N}-E_f|-\hbar\omega)\right].
\label{eq32}
\end{eqnarray} 
The delta function in the gain coefficient expression is modeled by a Lorentzian with the same FWHM as for electron-LO scattering rates.

\section {Numerical results and discussion}

\begin{figure}
\includegraphics[width=0.7\textwidth]{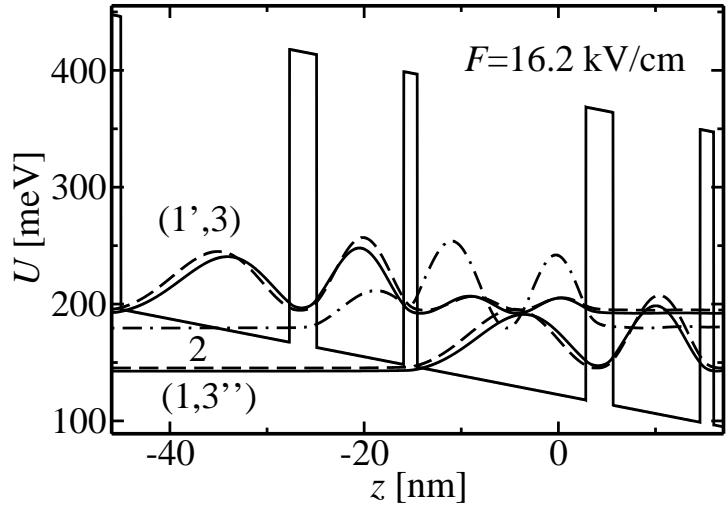}
\caption{A schematic diagram of the conduction band profile, size-quantized energy levels from which Landau levels originate and squared wave functions for one full period and parts of adjacent periods of the GaAs/AlGaAs QCL for zero magnetic field and an electric field of $16.2$~kV/cm. States $1$ and $1'$ (solid line), $2$ (dash-dotted line), $3$ and $3''$ (dashed line) denote the ground, lower laser and upper laser levels, respectively. State $1'$ belongs to the preceding period, while state $3''$ belongs to the following period.}
\label{fig1}
\end{figure}

As a prototypical system, we consider a QCL design which comprises a three-level scheme, and employs LO-phonon depopulation of the lower laser level to the ground state. No injector region is present and efficient injection into the upper laser level is enabled by its alignment with the ground level of the preceding period. The QCL period consists of two QW's (see Fig.~\ref{fig1}), one of which confines the ground and lower laser levels, whose energy difference is set to be approximately one LO phonon energy. The upper laser level is localized in the other well. This structure was chosen to be examined, instead of existing QCL's already investigated in the presence of a magnetic field,
due to its simplicity, and the dominant influence of electron-LO phonon interaction on the electron population dynamics, as will be explained in what follows.

The conduction band profile and electronic structure of the QCL in zero magnetic field and an electric field of $16.2$~kV/cm, is given in Fig.~\ref{fig1}. One QCL period includes a $2.8$~nm Al$_{0.3}$Ga$_{0.7}$As barrier, followed by a $9$~nm GaAs well, a $1.4$~nm Al$_{0.3}$Ga$_{0.7}$As barrier and a $17.4$~nm GaAs well. States $1$, $2$ and $3$ represent the ground, lower laser and upper laser levels, respectively, and states $1'$ and $3''$ represent the ground level of the preceding period and the upper laser level of the following period. The doping density was chosen to be $10^{11}$~cm$^{-2}$, in order to achieve relatively high gain in the THz range, while still having a small influence on the effective conduction band potential and making electron-electron processes less relevant. The temperature was set to $4$~K, since QCL's in a magnetic field are usually operated at low temperatures. The transition energy between the upper and lower laser levels is $15.2$~meV, and the energy difference between the ground state and the upper laser level of the following period is $2.6$~meV.

Relatively strong electron-electron scattering occurs between the ground state and the upper laser level of the next period, due to a large overlap and small energy difference. However, the semiclassical calculation performed for a similar structure showed that the electron-LO phonon scattering rates from the lower laser level to these levels, also relevant for the distribution of electrons between them, are considerably larger.~\cite{apl} Electron-electron processes between other states are less important, due to the large energy spacing of significantly populated LL's. Consequently, the population dynamics
are not significantly influenced by electron-electron scattering. We make an assumption that the same accounts for the polarization dynamics and neglect electron-electron interaction hereafter. 

\subsection{Electron populations}

\begin{figure}[htbp]
\includegraphics[width=0.7\textwidth]{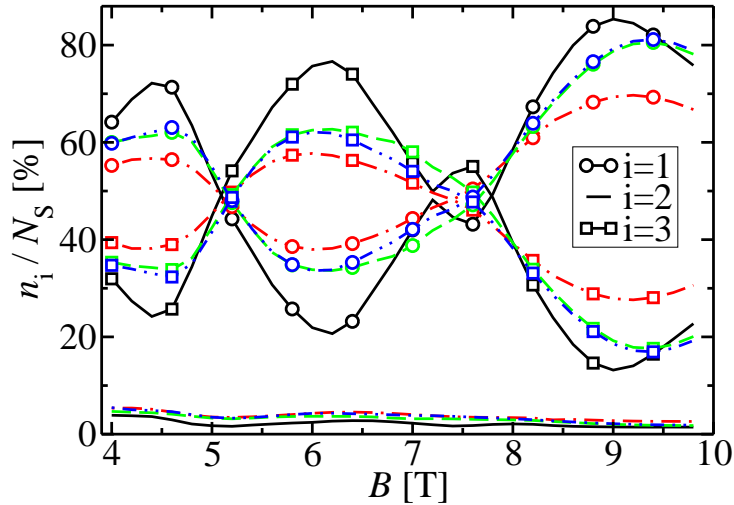}
\caption{(Color online). The electron population over QCL states (all Landau levels) vs magnetic field. States $1$, $2$, and $3$ represent the ground state, the lower laser level, and the upper laser level, respectively. $N_S$ is the total sheet density of electrons per period. Solid, dash-dotted, dashed and dash-double dotted lines represent non-Markovian ($\hbar\gamma_{ph}=1$~meV, and $\hbar\gamma_{ph}=2$~meV), Markovian ($\gamma=3$~meV) and Boltzmann ($\gamma=3$~meV) results, respectively.}
\label{fig2}
\end{figure}

The populations of all LL's associated with the ground state and the upper and lower laser levels, calculated using the non-Markovian, Markovian and Boltzmann model of electron transport, as functions of magnetic field, are shown in Fig.~\ref{fig2}. In the non-Markovian treatment, we used the values of the damping parameter of $\hbar\gamma_{ph}=1$~meV, and $\hbar\gamma_{ph}=2$~meV, while in the Markovian and Boltzmann description, the Lorentzian FWHM of $\gamma=3$~meV was taken. Regardless of the model used, their dependences on the magnetic field are generally similar. For certain magnetic fields ($4.6$~T, $9.2$~T), the energy difference between some LL's stemming from the upper laser level and the ground state becomes equal to one LO phonon energy, thus the electron-LO phonon interaction between them increases considerably. Consequently, the population of all LL's stemming from the upper laser level decreases reaching its minimum, while the opposite happens to the population of the LL's stemming from the ground state. 
Conversely, for intermediate magnetic fields ($6.2$~T), the population of the upper laser level is increased, while the ground state is depopulated. The population of the lower laser level practically does not change with magnetic field.

The populations obtained from the Markovian and Boltzmann description do not differ much, except in the range of magnetic fields between $6.2$~T and $8$~T, see Fig.~\ref{fig2}. In the fully nondiagonal Markovian (and non-Markovian) approach employed here, the coupling between populations and polarizations is accounted for, which results in the presence of phase coherence in the stationary state. In other words, during the time evolution of the system, scattering processes from populations to polarizations create a certain amount of polarization in the steady-state conditions. These and reversed processes (dephasing from polarizations to populations) may have an apparent impact on electron populations. More pronounced differences between the Markovian and Boltzmann populations, for magnetic fields between $6.2$~T and $8$~T, indicate a stronger interplay between populations and polarizations in the Markovian description and, hence, larger values of polarizations. Indeed, Figs.~\ref{fig3} and \ref{fig4} show that the polarizations between all LL's stemming from any two laser states are increased for these magnetic fields.

\begin{figure}[htbp]
\includegraphics[width=1.0\textwidth]{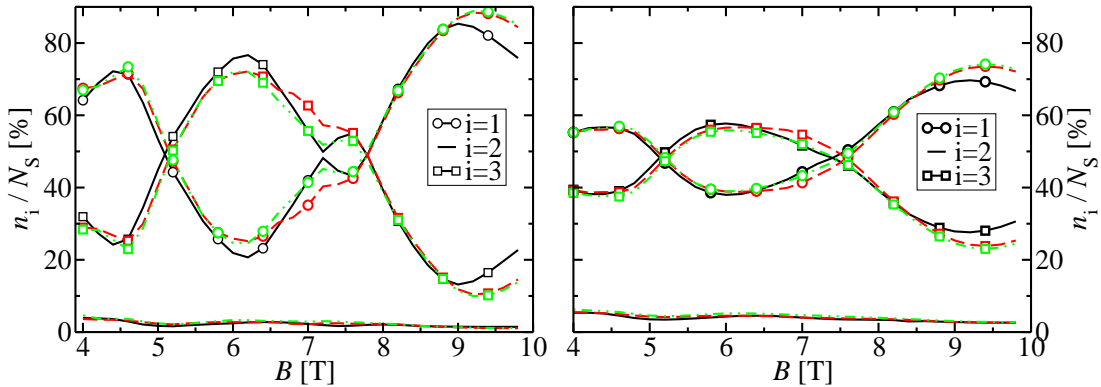}
\caption{(Color online). The electron population over QCL states (all Landau levels) vs magnetic field. States $1$, $2$, and $3$ represent the ground state, the lower laser level, and the upper laser level, respectively. $N_S$ is the total sheet density of electrons per period. Left: Solid, dashed and dash-double dotted lines represent non-Markovian ($\hbar\gamma_{ph}=1$~meV), Markovian ($\gamma=2$~meV) and Boltzmann ($\gamma=2$~meV) results, respectively. Right: Solid, dashed and dash-double dotted lines represent non-Markovian ($\hbar\gamma_{ph}=2$~meV), Markovian ($\gamma=4$~meV) and Boltzmann ($\gamma=4$~meV) results, respectively.}
\label{fig2a}
\end{figure}

On the other hand, the populations calculated from the quantum-kinetic model, for both values of the damping rate, show significant departure from the Markovian or Boltzmann results. For $\hbar\gamma_{ph}=1$~meV ($\hbar\gamma_{ph}=2$~meV), the amplitude of the oscillations of the upper laser level and the ground state populations is considerably larger (smaller) than in the Markovian or Boltzmann case, see Fig.~\ref{fig2}.
Such a drastic contrast between the results of the quantum-kinetic simulation for different damping values reflects the fact that these quantities actually describe collisional broadening,~\cite{dm} to which the LL populations are extremely sensitive. From that point of view, the difference in the population dynamics between the quantum-kinetic treatment and the Markovian or Boltzmann approach for $\gamma=3$~meV may be explained in terms of correspondence of $\hbar\gamma_{ph}=1$~meV ($\hbar\gamma_{ph}=2$~meV) to a smaller (larger) broadening, and hence, a smaller (larger) Lorentzian FWHM.
Indeed, the Markovian and Boltzmann calculation for different values of FWHM showed that the LL populations for $\gamma=2$~meV and $\gamma=4$~meV are very similar to the quantum-kinetic results for $\hbar\gamma_{ph}=1$~meV and $\hbar\gamma_{ph}=2$~meV, respectively, see Fig.~\ref{fig2a}. Nevertheless, this does not mean that there is one-to-one correspondence between these electron transport models for such values of FWHM and damping parameter. The non-Markovian description accounts for the memory of scattering and dephasing processes, which corresponds to a quantum-mechanical energy-time uncertainty. In comparison, the Markovian or Boltzmann dynamics take into account only energy-conserving processes, here, however, relaxed by the assumption that all LL's are broadened. This may lead to different values of polarizations, gain and current although the populations are quite similar. In order to give a fair comparison between different models presented here, we restrict our further analysis to a choice of the pairs of phenomenological parameters for which the populations are almost identical ($\gamma=2$~meV and $\hbar\gamma_{ph}=1$~meV; $\gamma=4$~meV and $\hbar\gamma_{ph}=2$~meV), and then we compare the results for other physical quantities.   

\subsection{Electron polarizations}

\begin{figure}[htbp]
\includegraphics[width=0.7\textwidth]{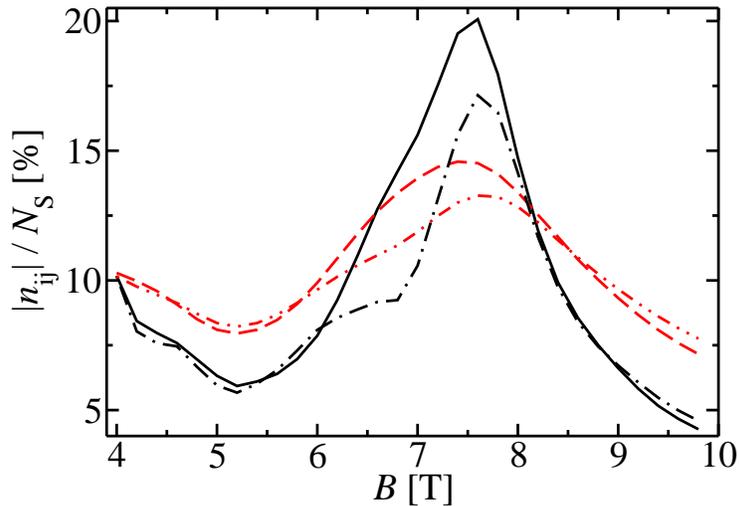}
\caption{(Color online). The electron polarization between the ground state of the preceding period and the upper laser level (all Landau levels) vs magnetic field. $N_S$ is the total sheet density of electrons per period. Solid, dashed, dash-dotted and dash-double dotted lines represent non-Markovian ($\hbar\gamma_{ph}=1$~meV and $\hbar\gamma_{ph}=2$~meV) and Markovian ($\gamma=2$~meV and $\gamma=4$~meV) results, respectively.}
\label{fig3}
\end{figure}

\begin{figure}[htbp]
\includegraphics[width=1.0\textwidth]{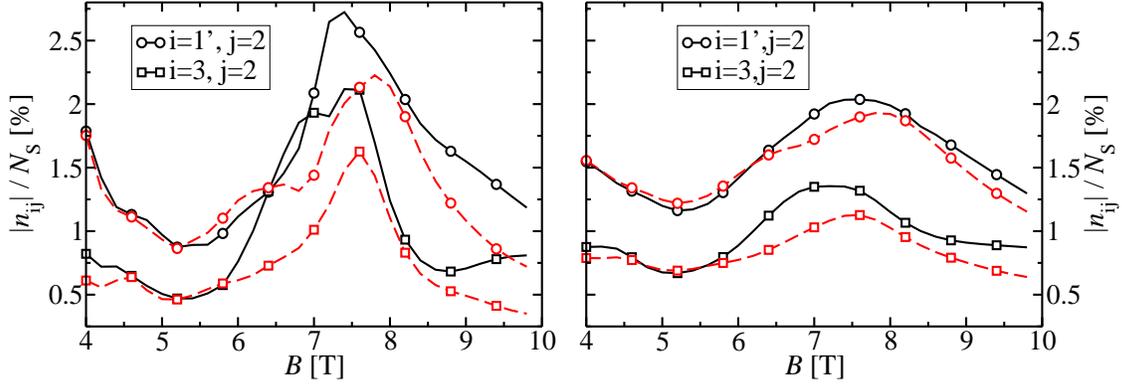}
\caption{(Color online). The electron polarization between the lower laser level and other QCL states (all Landau levels) vs magnetic field. States $1'$, $2$, and $3$ represent the ground state of the preceding period, the lower laser level, and the upper laser level, respectively. $N_S$ is the total sheet density of electrons per period. Left: Solid and dashed lines represent non-Markovian ($\hbar\gamma_{ph}=1$~meV) and Markovian ($\gamma=2$~meV) results, respectively. Right: Solid and dashed lines represent non-Markovian ($\hbar\gamma_{ph}=2$~meV) and Markovian ($\gamma=4$~meV) results, respectively.}
\label{fig4}
\end{figure}

Figs.~\ref{fig3} and \ref{fig4} show the most prominent polarizations between all LL's associated with pairs of laser states, obtained from the Markovian ($\gamma=2$~meV and $\gamma=4$~meV) and non-Markovian models ($\hbar\gamma_{ph}=1$~meV and $\hbar\gamma_{ph}=2$~meV), as they depend on the magnetic field. In all cases, the polarization between the ground state and the upper laser level of the next period, shown in Fig.~\ref{fig3}, is considerable ($\sim 10$~\%), due to the fact that these levels actually constitute a doublet state. Although their overlap does not change with magnetic field, their polarization does, since the LL electronic structure and all scattering/dephasing processes change as well. The polarizations between the upper laser level or the ground state of the previous period and the lower laser level are an order of magnitude smaller ($\sim 1$~\%), see Fig.~\ref{fig4}. This rapidly decreasing trend continues for the polarizations between other pairs of states, due to strong electron-LO phonon dephasing.
 
\begin{figure}[htbp]
\includegraphics[width=1.0\textwidth]{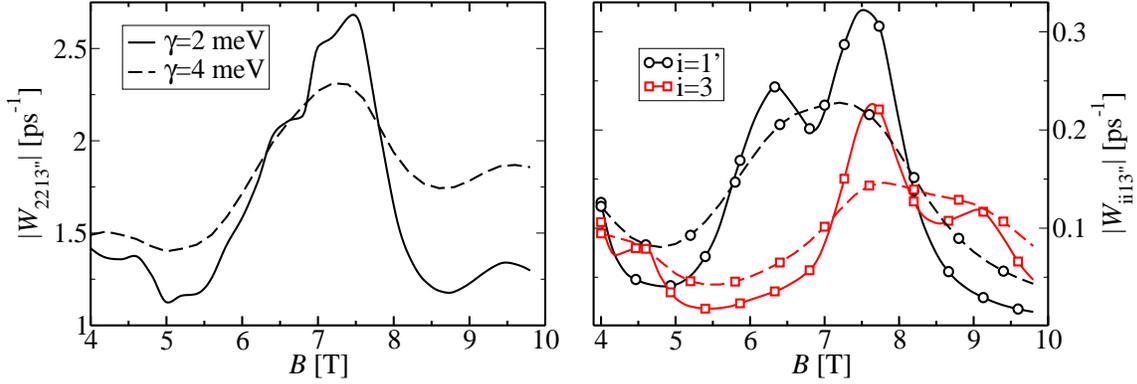}
\caption{(Color online). Average scattering rates (see text for explanation) from QCL states to polarizations between QCL states vs magnetic field, calculated in the Markovian approach for $\gamma=2$~meV (solid lines) and $\gamma=4$~meV (dashed lines). States $1$, $2$ and $3$ represent the ground state, the lower laser level and the upper laser level, respectively. States $1'$ and $3''$ denote the ground state of the preceding period, and the upper laser level of the following period.}
\label{fig4a}
\end{figure}

Generally, nondiagonal contribution curves in each Markovian case are offset to smaller values compared to the corresponding non-Markovian ones ($\gamma=2$~meV versus $\hbar\gamma_{ph}=1$~meV; $\gamma=4$~meV versus $\hbar\gamma_{ph}=2$~meV). 
This is caused by smaller scattering rates from populations to polarizations in the Markovian case, since they do not include the memory of the interaction. Deviations from this for some magnetic fields are a consequence of limited analogy between the two models for the broadening parameters $\gamma$ and $\gamma_{ph}$ used in the calculation. The coherences in the Markovian case for $\gamma=2$~meV have larger peaks than for $\gamma=4$~meV, while away from the peak, their values become smaller. The effect of a smaller FWHM is that the scattering rates responsible for the formation of polarizations have larger peak and lower valley values. The same conclusion applies to the non-Markovian results for different damping parameters. To illustrate this, the most influential scattering rates from populations to the polarization between the states of the doublet in the Markovian case versus magnetic field are shown in Fig.~\ref{fig4a}. The average scattering rate from the polarizations between LL's originating from state $m_{i_1}$ and $m_{i_2}$ into the polarizations between LL's originating from states $m_{i_4}$ and $m_{i_3}$ may be defined in the Markovian approach according to 
\begin{eqnarray}
W_{m_{i_1}m_{i_2}m_{i_3}m_{i_4}}^a=\sum_{m_{i_5}}\sum_{j_{i_1}-j_{i_5}}n_{\left|m_{i_1},j_{i_1}\right>\left|m_{i_5},j_{i_5}\right>}\Gamma_{\left|m_{i_2},j_{i_2}\right>,\left|m_{i_3},j_{i_3}\right>,\left|m_{i_4},j_{i_4}\right>,\left|m_{i_5},j_{i_5}\right>}^{out}\nonumber\\
(\delta_{\left|m_{i_4},j_{i_4}\right>,\left|m_{i_3},j_{i_3}\right>}-\alpha_Bn_{\left|m_{i_4},j_{i_4}\right>\left|m_{i_3},j_{i_3}\right>})/\sum_{m_{i_5}}\sum_{j_{i_1},j_{i_5}}n_{\left|m_{i_1},j_{i_1}\right>,\left|m_{i_5},j_{i_5}\right>}\nonumber\\
+\sum_{m_{i_5}}\sum_{j_{i_1}-j_{i_5}}n_{\left|m_{i_5},j_{i_5}\right>\left|m_{i_2},j_{i_2}\right>}\Gamma_{\left|m_{i_1},j_{i_1}\right>,\left|m_{i_4},j_{i_4}\right>,\left|m_{i_3},j_{i_3}\right>,\left|m_{i_5},j_{i_5}\right>}^{out\;*}\nonumber\\
(\delta_{\left|m_{i_4},j_{i_4}\right>,\left|m_{i_3},j_{i_3}\right>}-\alpha_Bn_{\left|m_{i_4},j_{i_4}\right>\left|m_{i_3},j_{i_3}\right>})/\sum_{m_{i_5}}\sum_{j_{i_1},j_{i_5}}n_{\left|m_{i_5},j_{i_5}\right>,\left|m_{i_2},j_{i_2}\right>}.
\label{wa}
\end{eqnarray}
Under the condition that $m_{i_1}=m_{i_2}$, 
Eq.~(\ref{wa}) gives the average scattering rate from LL's associated with state $m_{i_1}$ into the polarizations between LL's associated with from states $m_{i_4}$ and $m_{i_3}$. Also, the condition $m_{i_1}=m_{i_2}$ and $m_{i_3}=m_{i_4}$ gives the average scattering rate from LL's stemming from state $m_{i_1}$ into LL's stemming from state $m_{i_3}$, which is equivalent to the Boltzmann scattering rate.

\subsection{Optical gain}

\begin{figure}[htbp]
\includegraphics[width=1.0\textwidth]{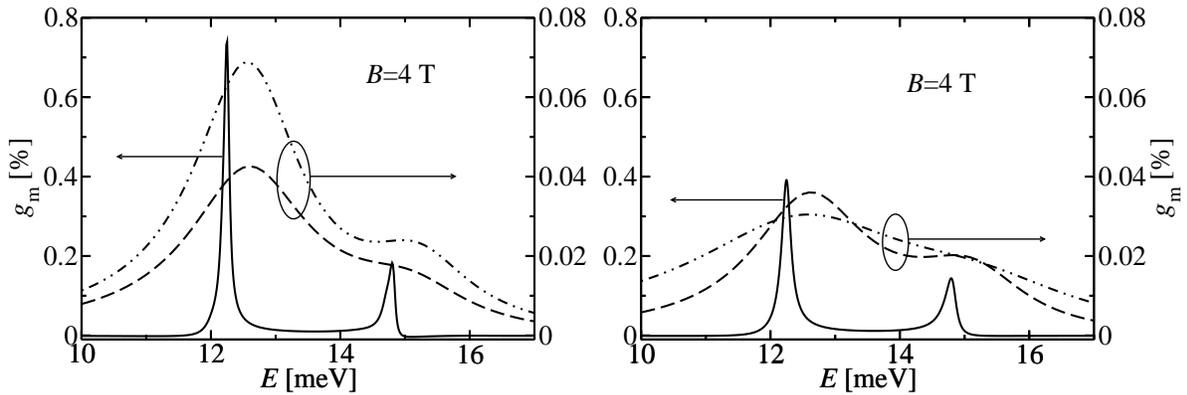}
\caption{Optical gain vs energy for a magnetic field of $4$~T. The energy range is in the vicinity of the optical transition energies. Left: Solid, dashed and dash-double dotted lines represent non-Markovian ($\hbar\gamma_{ph}=1$~meV), Markovian ($\gamma=2$~meV) and Boltzmann ($\gamma=2$~meV) results, respectively. Right: Solid, dashed and dash-double dotted lines represent non-Markovian ($\hbar\gamma_{ph}=2$~meV), Markovian ($\gamma=4$~meV) and Boltzmann ($\gamma=4$~meV) results, respectively.}
\label{fig5}
\end{figure}

\begin{figure}[htbp]
\includegraphics[width=1.0\textwidth]{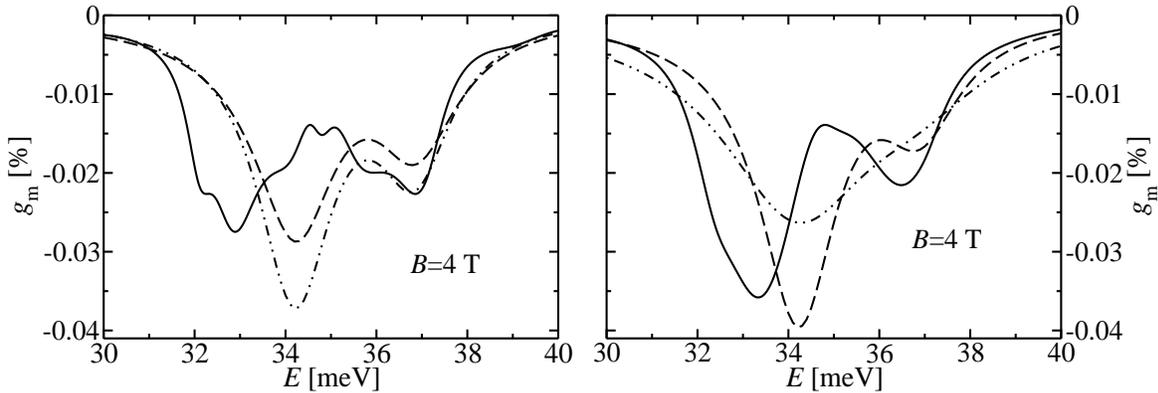}
\caption{Optical gain vs energy for a magnetic field of $4$~T. The energy range is in the vicinity of one longitudinal optical phonon energy. Left: Solid, dashed and dash-double dotted lines represent non-Markovian ($\hbar\gamma_{ph}=1$~meV), Markovian ($\gamma=2$~meV) and Boltzmann ($\gamma=2$~meV) results, respectively. Right: Solid, dashed and dash-double dotted lines represent non-Markovian ($\hbar\gamma_{ph}=2$~meV), Markovian ($\gamma=4$~meV) and Boltzmann ($\gamma=4$~meV) results, respectively.}
\label{fig6}
\end{figure}

The gain spectra for a magnetic field of $4$~T in the energy range close to the optical transition energies, and one LO phonon energy, are shown in Figs.~\ref{fig5} and \ref{fig6}, respectively. The gain was calculated for the non-Markovian ($\hbar\gamma_{ph}=1$~meV and $\hbar\gamma_{ph}=2$~meV), Markovian and Boltzmann ($\gamma=2$~meV and $\gamma=4$~meV) dynamics. 
The modal gain, defined as $g_m(\omega)=g(\omega)d$, and obtained from the Boltzmann theory, has identical major features as the one predicted from the Markovian approach, for the same value of FWHM. Disregarding nondiagonal scattering and dephasing processes in the Boltzmann model, which affect the gain linewidth, results only in a quantitative modification of the Markovian prediction.

In the case of the non-Markovian dynamics, the gain linewidth is significantly decreased for optical transition energies in comparison to the corresponding Markovian and Boltzmann estimates ($\approx 10$ times). This result might seem somewhat unexpected because scattering and dephasing are increased compared to the Markovian treatment by including the memory of the interaction process. 
However, in the non-Markovian approach, the broadening caused by the scattering terms is energy dependent.~\cite{absspectra} 
The frequency dependence of the quantity $\Delta K_{i_1i_2i_3i_4}(\omega)$ (see Eq.~(\ref{eq29})), correlated to phonon-assisted matrices, shows that the non-Markovian calculation does not yield energy conservation
in the scattering processes, but includes an additional $\hbar\omega$ contribution. Then, in the coefficient associated with $\Delta n_{\left|2,j\right>\left|3,j\right>}(\omega)$ in Eq.~(\ref{eq29}), if $\hbar\omega\approx \bar{E}_3-\bar{E}_2$, the sum of the terms reciprocal to $E_{i_3}+\hbar\omega_{\text{LO}}-E_{\left|2,j\right>}-\hbar\omega-i\hbar\gamma_{ph}\approx E_{i_3}-E_{\left|2,j\right>}+23.65$~meV$-i\hbar\gamma_{ph}$ and $E_{\left|3,j\right>}+\hbar\omega_{\text{LO}}-E_{i_3}-\hbar\omega-i\hbar\gamma_{ph}\approx E_{\left|3,j\right>}-E_{i_3}+23.65$~meV$-i\hbar\gamma_{ph}$ appear. For the electronic structure of the QCL considered, these terms are not resonant for the majority of $i_3=1,...,2N$, and their contribution to the coefficient related to $\Delta n_{\left|2,j\right>\left|3,j\right>}(\omega)$ is small. Hence, $\Delta n_{\left|2,j\right>\left|3,j\right>}(\omega)$ is large, and so is the gain. In contrast, the same coefficient, but in the Markovian description, contains the sum of the terms proportional to $\delta(E_{i_5}+\hbar\omega_{\text{LO}}-E_{i_4})$ and $\delta(E_{i_5}-\hbar\omega_{\text{LO}}-E_{i_4})$ (see Eq.~(\ref{eq30})), which are always resonant for $i_4=\left|1,j\right>$, or $i_4=\left|3'',j\right>$, and $i_5=\left|2,j\right>$ (or vice versa). Therefore, the coefficient associated with $\Delta n_{\left|2,j\right>\left|3,j\right>}(\omega)$ is large, and $\Delta n_{\left|2,j\right>\left|3,j\right>}(\omega)$ and the gain are small. Similar arguments may be used to explain an increased gain peak and narrow linewidth in the non-Markovian treatment if $\hbar\omega\approx \bar{E}_{1'}-\bar{E}_2$. However, in a real QCL device, it is likely that interaction of electrons with impurities, interface defects and other electrons, as well as imperfect periodicity, will broaden LL's making the gain linewidth for optical transition energies not as narrow as the non-Markovian model predicts. 
On the other hand, for $\hbar\omega\approx \bar{E}_2-\bar{E}_1\approx \hbar\omega_{\text{LO}}$ in the non-Markovian treatment, the terms inverse to $E_{i_3}+\hbar\omega_{\text{LO}}-E_{\left|1,j\right>}-\hbar\omega-i\hbar\gamma_{ph}\approx E_{i_3}-E_{\left|1,j\right>}-i\hbar\gamma_{ph}$ and $E_{\left|2,j\right>}+\hbar\omega_{\text{LO}}-E_{i_3}-\hbar\omega-i\hbar\gamma_{ph}\approx E_{\left|2,j\right>}-E_{i_3}-i\hbar\gamma_{ph}$ contribute to the coefficient related to $\Delta n_{\left|1,j\right>\left|2,j\right>}(\omega)$. Obviously, these terms are resonant for $i_3=\left|1,j\right>$ or $i_3=\left|2,j\right>$, thus $\Delta n_{\left|1,j\right>\left|2,j\right>}(\omega)$ and the gain are small and comparable to the Markovian case. The same situation occurs for $\hbar\omega\approx \bar{E}_2-\bar{E}_{3''}$.

In the Markovian limit, energy renormalizations, describing the polaron corrections to the band structure, are ignored. However, the polaron shift is always included in the quantum-kinetic treatment. It is more prominent for the energy transitions close to one LO phonon energy ($\sim 1$~meV), but it is also present for the optical transition energies ($\sim 0.4$~meV). 

\begin{figure}[htbp]
\includegraphics[width=1.0\textwidth]{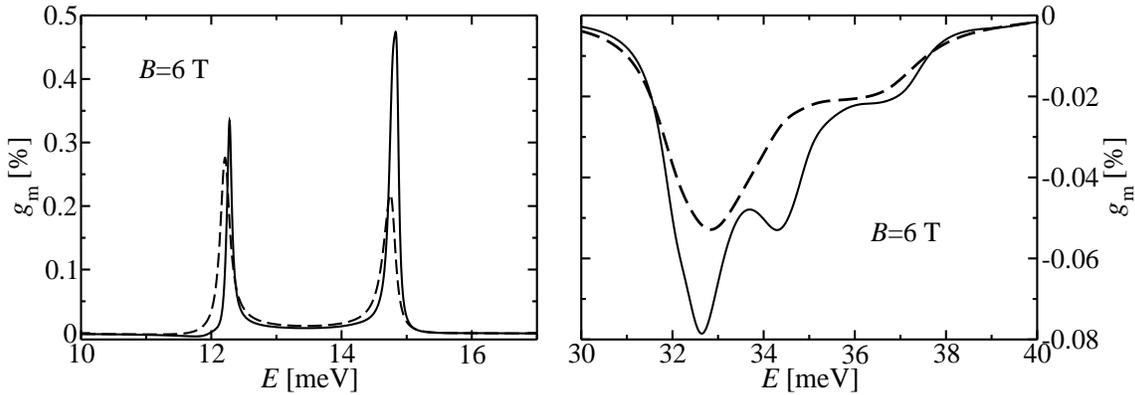}
\caption{Optical gain vs energy for a magnetic field of $6$~T calculated using the non-Markovian approach for $\hbar\gamma_{ph}=1$~meV (solid line) and $\hbar\gamma_{ph}=2$~meV (dashed line). Left: The energy range is in the vicinity of the optical transition energies. Right: The energy range is in the vicinity of one longitudinal optical phonon energy.}
\label{fig6b}
\end{figure}

Fig.~\ref{fig6b} illustrates the gain profile for a magnetic field of $6$~T in the energy range around the optical transition energies and one LO phonon energy, calculated from the non-Markovian model for $\hbar\gamma_{ph}=1$~meV and $\hbar\gamma_{ph}=2$~meV. 
Comparison of the non-Markovian gain spectra for different values of the damping parameter (see also Figs.~\ref{fig5} and \ref{fig6}) reveals pronounced differences between both the gain linewidth and peak values. Sensitivity of the results to the values of phenomenological parameters confirms the need for a self-consistent incorporation of higher order correlations in the quantum-kinetic model, which would require a significant increase in computational time.
 Moreover, for a magnetic field of $6$~T, apart from the two expected peaks associated with the transitions from the lower laser level to the ground state or the upper laser level of the subsequent period, an extra peak appears if $\hbar\gamma_{ph}=1$~meV is taken. Careful inspection of Eq.~(\ref{eq29}) reveals the presence of more resonances at the energies of $\hbar\omega\approx \pm(E_{\left|3'',j\right>}-E_{\left|1,j\right>})+\hbar\omega_{\text{LO}}=\pm 2.6$~meV$+\hbar\omega_{\text{LO}}$. At first, it may appear that one of these polaron resonances, related to the transition between  the ground state and the upper laser level of the subsequent period, is manifested via that additional peak in the gain/absorption spectra (so-called polaron satellite).~\cite{stephan} However, this is not entirely the case here. Despite those resonant terms, the coefficient associated with $\Delta n_{\left|3'',j\right>\left|1,j\right>}(\omega)$ is large, since $E_{\left|1,j\right>}-E_{\left|3'',j\right>}-\hbar\omega$ is large (see Eq.~(\ref{eq29})).
 At the same time, the coefficients associated with $\Delta n_{\left|1,j\right>\left|2,j\right>}(\omega)$ and $\Delta n_{\left|3'',j\right>\left|2,j\right>}(\omega)$ are smaller, because $E_{\left|2,j\right>}-E_{\left|1,j\right>}-\hbar\omega$ and $E_{\left|2,j\right>}-E_{\left|3'',j\right>}-\hbar\omega$ are small, and, additionally, they also include the aforementioned polaron resonant terms, as well as the terms reciprocal to $E_{i_3}+\hbar\omega_{\text{LO}}-E_{\left|1,j\right>}-\hbar\omega-i\hbar\gamma_{ph}\approx E_{i_3}-E_{\left|1,j\right>}-i\hbar\gamma_{ph}$ or $E_{\left|2,j\right>}+\hbar\omega_{\text{LO}}-E_{i_3}-\hbar\omega-i\hbar\gamma_{ph}\approx E_{\left|2,j\right>}-E_{i_3}-i\hbar\gamma_{ph}$, resonant for $i_3=\left|1,j\right>$ or $i_3=\left|2,j\right>$. As a consequence, $\Delta n_{\left|1,j\right>\left|2,j\right>}(\omega)$ or $\Delta n_{\left|3'',j\right>\left|2,j\right>}(\omega)$ constitute much larger fraction of the total gain than $\Delta n_{\left|3'',j\right>\left|1,j\right>}(\omega)$, unlike it is the case in Ref.~\onlinecite{stephan}, where only two subbands were considered. Here, the additional peak for $B=6$~T and $\hbar\gamma_{ph}=1$~meV is actually related to the transition between the lower laser level and the upper laser level of the next period (the calculation showed that $\Delta n_{\left|1,j\right>\left|2,j\right>}(\omega)<\Delta n_{\left|3'',j\right>\left|2,j\right>}(\omega)$), as well as the peak at $\hbar\omega=32.7$~meV which is also present for $\hbar\gamma_{ph}=2$~meV, while the peak at $\hbar\omega=36.8$~meV is associated to the transition between the lower laser level and the ground state. In fact, the previous analysis of the terms which are resonant for the energies in the vicinity of one LO-phonon energy suggests that up to $6$ peaks could emerge in the absorption spectra in that energy range. Due to the their proximity, some of them may be obscured by other stronger peaks for different values of the damping parameter and magnetic fields. In Fig.~\ref{fig6}, for the value of the damping parameter of $\hbar\gamma_{ph}=1$~meV, we can notice that, in addition to $2$ well defined peaks, $4$ small ones also appear.

\begin{figure}[htbp]
\includegraphics[width=1.0\textwidth]{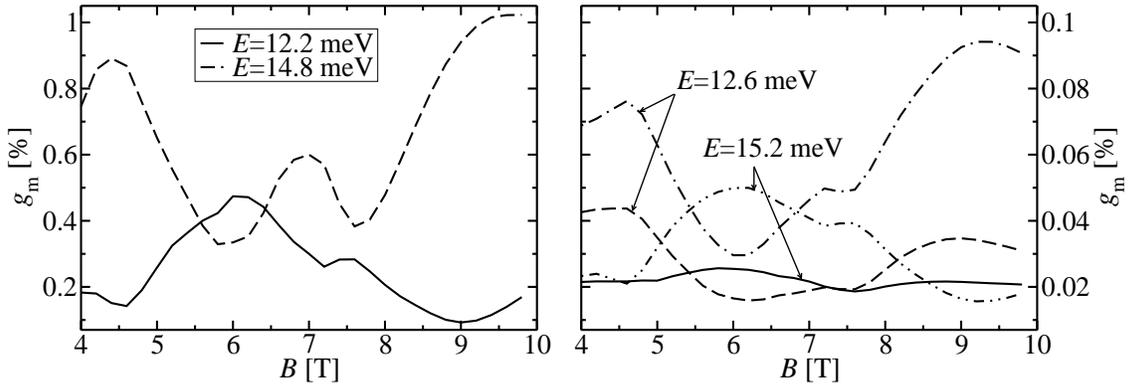}
\caption{Maximal gain vs magnetic field for transitions from the upper laser level and the ground state of the preceding period to the lower laser level. Left: The non-Markovian ($\hbar\gamma_{ph}=1$~meV) results are represented by solid and dashed lines, respectively. Right: The Markovian ($\gamma=2$~meV) results are represented by solid and dashed lines, respectively. The Boltzmann ($\gamma=2$~meV) results are represented by dash-dotted and dash-double dotted lines, respectively.} 
\label{fig6c}
\end{figure}

Fig.~\ref{fig6c} shows peak gain values versus magnetic field dependence, for transitions from the upper laser level and the ground level of the preceding period to the lower laser level, obtained from the non-Markovian ($\hbar\gamma_{ph}=1$~meV), Markovian and Boltzmann ($\gamma=2$~meV) approach. 
The general trend of the non-Markovian gain is fairly similar to the Markovian or Boltzmann results, although the former has larger values.
The oscillations of the gain in both types of transitions reproduce reasonably well the oscillations of the related populations (see Fig.~\ref{fig2a}). It should be noted that, for most magnetic fields, the dominant gain spectral component is not generated in the transitions from the upper laser level, but from the ground state of the preceding period, even at some magnetic fields for which the upper laser level is more populated. The reason for that is a slightly larger dipole matrix element between the ground state of the preceding period and the lower laser level.

\subsection{Current}

\begin{figure}[htbp]
\includegraphics[width=1.\textwidth]{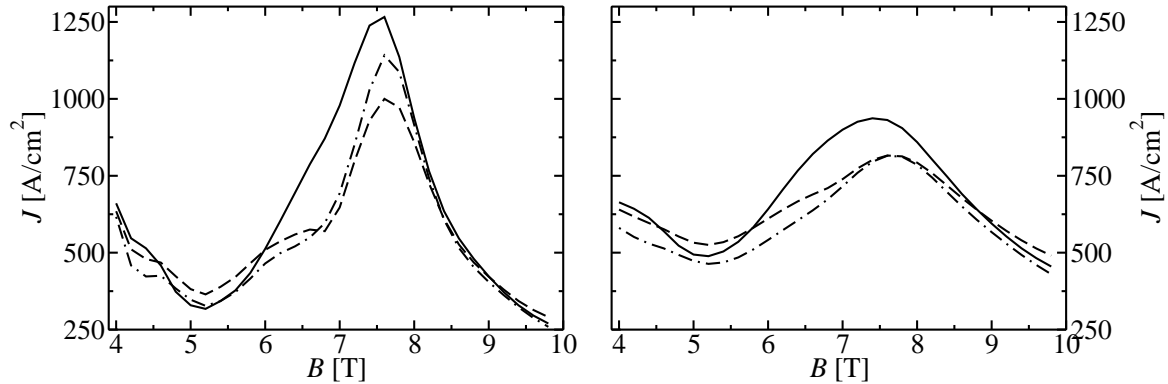}
\caption{Left: Current density vs magnetic field dependence. Left: Solid, dashed and dash-dotted lines represent non-Markovian ($\hbar\gamma_{ph}=1$~meV), Markovian ($\gamma=2$~meV) and Boltzmann ($\gamma=2$~meV) results, respectively. Right: Solid, dashed and dash-dotted lines represent non-Markovian ($\hbar\gamma_{ph}=2$~meV), Markovian ($\gamma=4$~meV) and Boltzmann ($\gamma=4$~meV) results, respectively.}
\label{fig7}
\end{figure}

The current densities as functions of magnetic field, calculated using the non-Markovian ($\hbar\gamma_{ph}=1$~meV and $\hbar\gamma_{ph}=2$~meV), Markovian and Boltzmann description ($\gamma=2$~meV and $\gamma=4$~meV), are shown in Fig.~\ref{fig7}. From Eq.~(\ref{eq23}), used in the Markovian and non-Markovian approach, it follows that diagonal density matrix elements do not contribute to the total current.~\cite{iotti3,coherent} Therefore, the electron transport is entirely due to nondiagonal density matrix contributions i.e. scattering induced phase coherences between the laser states. This quantum-mechanical picture of completely coherent current is in a stark contrast with the semiclassical picture of transport through scattering transitions. However, both descriptions give similar results, see Fig.~\ref{fig7}. Here, the nondiagonal density matrix element are considerably smaller in comparison to the diagonal ones (see Figs.~\ref{fig2a}, \ref{fig3} and \ref{fig4}), thus the former may be approximated in terms of the latter,~\cite{jpc,coherent} giving Eq.~(\ref{eq24}) used in the calculation of the semiclassical current, and resulting in comparable values of the current density.

\begin{figure}[htbp]
\includegraphics[width=0.7\textwidth]{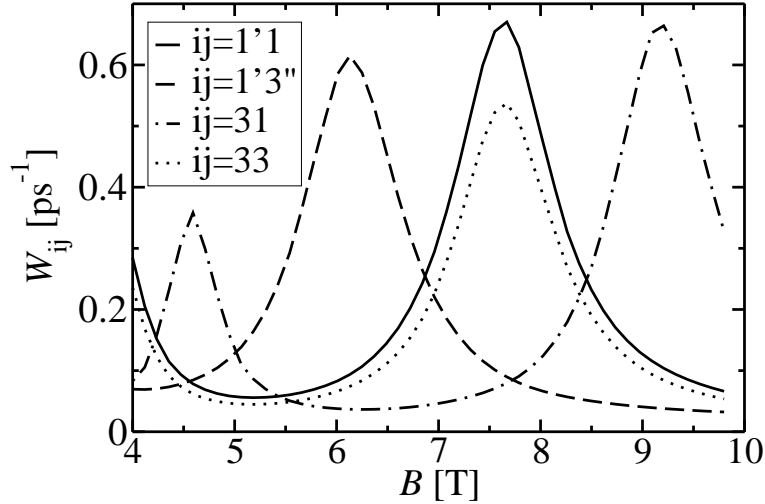}
\caption{Average scattering rates (see text for explanation) between QCL states vs magnetic field, calculated in the Boltzmann description for $\gamma=2$~meV. States $1$, $2$ and $3$ represent the ground state, the lower laser level and the upper laser level, respectively. States $1'$ and $3''$ denote the ground state of the preceding period, and the upper laser level of the following period.}
\label{fig8}
\end{figure}

In the semiclassical interpretation, the electron transport channel from one doublet directly to the subsequent doublet is as important as the channel which additionally involves the lower laser level. The scattering rates from the lower laser level to the ground state and the upper laser level of the next period do not exhibit pronounced oscillations with magnetic field since those energy transitions are close to one LO phonon energy. Therefore, the current versus magnetic field dependence in the semiclassical picture is mainly determined by the scattering rates between the doublet states, shown in Fig.~\ref{fig8}, and their populations. In the Markovian and non-Markovian description, the current is completely determined by the polarizations between the QCL states, see the diagrams shown in Figs.~\ref{fig3}, \ref{fig4} and \ref{fig7}. In both approaches, the current density curves reproduce well the main features of the related polarization curves. Also, the discrepancies between the current densities estimated from the Markovian and non-Markovian treatments are identical to those between the related polarizations.

\section {Conclusion}

We presented a quantum kinetic description of electron dynamics and gain in QCL's subjected to a magnetic field, based on the density matrix formalism. As a first step, electron-LO phonon interaction was considered as the most relevant scattering/dephasing mechanism. Nonequilibrium stationary state populations and polarizations for an example QCL structure were calculated using various kinetic models (non-Markovian, Markovian and Boltzmann). Since all of these models contain a phenomenological parameter, we chose the sets of those parameters so that steady-state populations are similar for a range of magnetic fields, and then we compared other relevant quantities (polarizations, gain, current). In both the Markovian and non-Markovian approach, coherent polarizations induced by electron-LO phonon interaction were found to be relatively small. 
This in turn led to similar values of the entirely coherent current in the Markovian and non-Markovian picture compared to the values of entirely incoherent current in the Boltzmann interpretation. Gain spectra in the non-Markovian treatment showed considerably narrow linewidths for optical transitions and evidence of polaron formation, in contrast to the Markovian and Boltzmann predictions.


\appendix
\section{}

This appendix presents a detailed derivation of the quantum kinetic equations for electron populations and polarizations in QW's under an applied magnetic field. The procedure for obtaining appropriate expressions in the Markovian approximation is also given.

Eqs.~(\ref{eq11}) and (\ref{eq15}) represent a starting point for the derivation of the quantum kinetic equations of motion, giving the following expressions
\begin{eqnarray}
\frac{d}{dt}\delta s_{k,{\bf q},k-q_y}^{i_1i_2}=\frac{1}{i\hbar}(E_{i_2}+\hbar\omega_{\text{LO}}-E_{i_1})\delta s_{k,{\bf q},k-q_y}^{i_1i_2}-\gamma_{ph}\delta s_{k,{\bf q},k-q_y}^{i_1i_2}+\frac{1}{i\hbar}\sum_{i_4,i_5}g_{\bf q}^*H_{j_{i_5}j_{i_4}}^*(k,k-q_y,q_x)\nonumber\\G_{m_{i_5}m_{i_4}}^*(q_z)\left[(n_0+1)f_{i_1i_5,k}(\delta_{i_4,i_2}-f_{i_4i_2,k-q_y})-n_0f_{i_4i_2,k-q_y}(\delta_{i_1,i_5}-f_{i_1i_5,k})\right],\nonumber\\
\frac{d}{dt}f_{i_1i_2,k}\left|_{\hat{H}_{ep}}\right.=\frac{1}{i\hbar}\sum_{i_3,{\bf q}}\left(g_{\bf q}H_{j_{i_2}j_{i_3}}(k,k-q_y,q_x)G_{m_{i_2}m_{i_3}}(q_z)\delta s_{k,{\bf q},k-q_y}^{i_1i_3}\right.\nonumber\\\left.+g_{\bf q}^*H_{j_{i_3}j_{i_2}}^*(k+q_y,k,q_x)G_{m_{i_3}m_{i_2}}^*(q_z)\delta s_{k+q_y,{\bf q},k}^{i_3i_1*}-g_{\bf q}H_{j_{i_3}j_{i_1}}(k+q_y,k,q_x)G_{m_{i_3}m_{i_1}}(q_z)\delta s_{k+q_y,{\bf q},k}^{i_3i_2}\right.\nonumber\\\left.-g_{\bf q}^*H_{j_{i_1}j_{i_3}}^*(k,k-q_y,q_x)G_{m_{i_1}m_{i_3}}^*(q_z)\delta s_{k,{\bf q},k-q_y}^{i_2i_3*}\right).
\label{eq33}
\end{eqnarray}
The assumption of initially uncorrelated system ($\lim_{t\rightarrow -\infty}\delta s_{k,{\bf q},k-q_y}^{i_1i_2}(t)=0$) gives
\begin{eqnarray}
\delta s_{k,{\bf q},k-q_y}^{i_1i_2}=\frac{1}{i\hbar}\sum_{i_4,i_5}g_{\bf q}^*H_{j_{i_5}j_{i_4}}^*(k,k-q_y,q_x)G_{m_{i_5}m_{i_4}}^*(q_z)\nonumber\\\int_{-\infty}^t dt'\;\exp\left\{\left[\frac{1}{i\hbar}(E_{i_2}+\hbar\omega_{\text{LO}}-E_{i_1})-\gamma_{ph}\right](t-t')\right\}\left[(n_0+1)f_{i_1i_5,k}(\delta_{i_4,i_2}-f_{i_4i_2,k-q_y})\right.\nonumber\\\left.-n_0f_{i_4i_2,k-q_y}(\delta_{i_1,i_5}-f_{i_1i_5,k})\right]\nonumber\\=\sum_{i_4,i_5}g_{\bf q}^*H_{j_{i_5}j_{i_4}}^*(k,k-q_y,q_x)G_{m_{i_5}m_{i_4}}^*(q_z)\delta K_{k,{\bf q},k-q_y}^{i_1i_2i_4i_5},
\label{eq34}
\end{eqnarray}
where the quantities $\delta K_{k,{\bf q},k-q_y}^{i_1i_2i_4i_5}$ related to the phonon assisted matrices $\delta s_{k,{\bf q},k-q_y}^{i_1i_2}$ are represented as
\begin{eqnarray}
\frac{d}{dt}\delta K_{k,{\bf q},k-q_y}^{i_1i_2i_4i_5}=\frac{1}{i\hbar}(E_{i_2}+\hbar\omega_{\text{LO}}-E_{i_1})\delta K_{k,{\bf q},k-q_y}^{i_1i_2i_4i_5}-\gamma_{ph}\delta K_{k,{\bf q},k-q_y}^{i_1i_2i_4i_5}\nonumber\\+\frac{1}{i\hbar}\left[(n_0+1)f_{i_1i_5,k}(\delta_{i_4,i_2}-f_{i_4i_2,k-q_y})-n_0f_{i_4i_2,k-q_y}(\delta_{i_1,i_5}-f_{i_1i_5,k})\right].
\label{eq35}
\end{eqnarray}
Simple algebra then leads to the equations for the dynamics of populations and polarizations in the form
\begin{eqnarray}
\frac{d}{dt}f_{i_1i_2,k}\left|_{\hat{H}_{ep}}\right.=\frac{1}{i\hbar}\sum_{i_3,i_4,i_5,{\bf q}}|g_{\bf q}|^2\nonumber\\\left(H_{j_{i_2}j_{i_3}}(k,k-q_y,q_x)H_{j_{i_5}j_{i_4}}^*(k,k-q_y,q_x)G_{m_{i_2}m_{i_3}}(q_z)G_{m_{i_5}m_{i_4}}(q_z)^*\delta K_{k,{\bf q},k-q_y}^{i_1i_3i_4i_5}\right.\nonumber\\\left.+H_{j_{i_3}j_{i_2}}^*(k+q_y,k,q_x)H_{j_{i_4}j_{i_5}}(k+q_y,k,q_x)G_{m_{i_3}m_{i_2}}^*(q_z)G_{m_{i_4}m_{i_5}}(q_z)\delta K_{k+q_y,{\bf q},k}^{i_3i_1i_5i_4*}\right.\nonumber\\\left.-H_{j_{i_3}j_{i_1}}(k+q_y,k,q_x)H_{j_{i_4}j_{i_5}}^*(k+q_y,k,q_x)G_{m_{i_3}m_{i_1}}(q_z)G_{m_{i_4}m_{i_5}}^*(q_z)\delta K_{k+q_y,{\bf q},k}^{i_3i_2i_5i_4}\right.\nonumber\\\left.-H_{j_{i_1}j_{i_3}}^*(k,k-q_y,q_x)H_{j_{i_5}j_{i_4}}(k,k-q_y,q_x)G_{m_{i_1}m_{i_3}}^*(q_z)G_{m_{i_5}m_{i_4}}(q_z)\delta K_{k,{\bf q},k-q_y}^{i_2i_3i_4i_5*}\right).
\label{eq36}
\end{eqnarray}

From the expression for the lateral overlap integrals derived in Appendix B (Eqs.~(\ref{eq48}) and (\ref{eq50})), it can be shown that the following expressions
\begin{eqnarray}
H_{j_{i_1}j_{i_3}}(k,k-q_y,q_x)H_{j_{i_5}j_{i_4}}^*(k,k-q_y,q_x)=|H_{j_{i_1}j_{i_3}}(q_{xy})||H_{j_{i_5}j_{i_4}}(q_{xy})|\nonumber\\e^{i\frac{\pi}{2}a_1}e^{i\theta(j_{i_1}+j_{i_4}-j_{i_3}-j_{i_5})}\nonumber\\
H_{j_{i_3}j_{i_1}}^*(k+q_y,k,q_x)H_{j_{i_4}j_{i_5}}(k+q_y,k,q_x)=|H_{j_{i_3}j_{i_1}}(q_{xy})||H_{j_{i_4}j_{i_5}}(q_{xy})|\nonumber\\e^{i\frac{\pi}{2}a_1}e^{i\theta(j_{i_1}+j_{i_4}-j_{i_3}-j_{i_5})},
\label{eq37}
\end{eqnarray}
where $\theta=\arg(q_x+iq_y)$ and
\begin{eqnarray}
a_1=\left\{\begin{array}{ll}
j_{i_1}+j_{i_4}-j_{i_3}-j_{i_5}, & j_{i_1}>j_{i_3} \land j_{i_5}>j_{i_4}\\
-j_{i_1}-j_{i_4}+j_{i_3}+j_{i_5}, & j_{i_1}<j_{i_3} \land j_{i_5}<j_{i_4}\\
j_{i_1}-j_{i_4}-j_{i_3}+j_{i_5}, & j_{i_1}>j_{i_3} \land j_{i_5}<j_{i_4}\\
-j_{i_1}+j_{i_4}+j_{i_3}-j_{i_5}, & j_{i_1}<j_{i_3} \land j_{i_5}>j_{i_4}\\
\end{array}
,\right.\nonumber
\end{eqnarray}
hold. Then, the temporal evolution of the density matrix elements read
\begin{eqnarray}
\frac{d}{dt}f_{i_1i_2,k}\left|_{\hat{H}_{ep}}\right.=\frac{1}{i\hbar}\sum_{i_3,i_4,i_5,{\bf q}}|g_{\bf q}|^2\nonumber\\\left(|H_{j_{i_2}j_{i_3}}(q_{xy})||H_{j_{i_5}j_{i_4}}(q_{xy})|G_{m_{i_2}m_{i_3}}(q_z)G_{m_{i_5}m_{i_4}}(q_z)^*e^{i\frac{\pi}{2}a_2}e^{i\theta(j_{i_2}+j_{i_4}-j_{i_3}-j_{i_5})}\delta K_{k,{\bf q},k-q_y}^{i_1i_3i_4i_5}\right.\nonumber\\\left.+|H_{j_{i_3}j_{i_2}}(q_{xy})||H_{j_{i_4}j_{i_5}}(q_{xy})|G_{m_{i_3}m_{i_2}}^*(q_z)G_{m_{i_4}m_{i_5}}(q_z)e^{i\frac{\pi}{2}a_2}e^{i\theta(j_{i_2}+j_{i_4}-j_{i_3}-j_{i_5})}\delta K_{k+q_y,{\bf q},k}^{i_3i_1i_5i_4*}\right.\nonumber\\\left.-|H_{j_{i_3}j_{i_1}}(q_{xy})||H_{j_{i_4}j_{i_5}}(q_{xy})|G_{m_{i_3}m_{i_1}}(q_z)G_{m_{i_4}m_{i_5}}^*(q_z)e^{-i\frac{\pi}{2}a_1}e^{-i\theta(j_{i_1}+j_{i_4}-j_{i_3}-j_{i_5})}\delta K_{k+q_y,{\bf q},k}^{i_3i_2i_5i_4}\right.\nonumber\\\left.-|H_{j_{i_1}j_{i_3}}(q_{xy})||H_{j_{i_5}j_{i_4}}(q_{xy})|G_{m_{i_1}m_{i_3}}^*(q_z)G_{m_{i_5}m_{i_4}}(q_z)e^{-i\frac{\pi}{2}a_1}e^{-i\theta(j_{i_1}+j_{i_4}-j_{i_3}-j_{i_5})}\delta K_{k,{\bf q},k-q_y}^{i_2i_3i_4i_5*}\right),
\label{eq38}
\end{eqnarray}
where the expression for $a_2$ is similar to $a_1$, with the exception that it contains index $j_{i_2}$ instead of $j_{i_1}$.
 
Close inspection of Eqs.~(\ref{eq35}) and (\ref{eq38}) shows that they represent independent identical subsystems of equations for each wave vector $k$, 
therefore the density matrix elements $f_{i_1i_2,k}$ and the quantities associated to phonon assisted matrices $\delta K_{k,{\bf q},k-q_y}^{i_1i_2i_4i_5}$ do not depend on the wave vector $k$ under the influence of electron-LO phonon interaction alone. It is obvious from Eqs.~(\ref{eq12}) and (\ref{eq13}) that the effect of free electrons system and interaction with light on the density matrix elements is identical. Furthermore, since the thermal equilibrium of phonon is assumed, phonon populations do not depend on the wave vector ${\bf q}$, and so does not $\delta K_{k,{\bf q},k-q_y}^{i_1i_2i_4i_5}$ ($\delta K_{k,{\bf q},k-q_y}^{i_1i_2i_4i_5}\rightarrow \delta K_{i_1i_2i_4i_5}$). Consequently, we have
\begin{eqnarray}
f_{i_1i_2,k}=\alpha_Bn_{i_1i_2},\;\alpha_B=\frac{\pi\hbar}{eB},\;n_{i_1i_2}=\frac{\sum_{k'}f_{i_1i_2,k'}}{L_xL_y},\nonumber\\
\frac{d}{dt}n_{i_1i_2}\left|_{\hat{H}_{ep}}\right.=\frac{1}{i\hbar}\sum_{i_3,i_4,i_5,{\bf q}}|g_{\bf q}|^2\nonumber\\\left(|H_{j_{i_2}j_{i_3}}(q_{xy})||H_{j_{i_5}j_{i_4}}(q_{xy})|G_{m_{i_2}m_{i_3}}(q_z)G_{m_{i_5}m_{i_4}}(q_z)^*e^{i\frac{\pi}{2}a_2}e^{i\theta(j_{i_2}+j_{i_4}-j_{i_3}-j_{i_5})}\delta K_{i_1i_3i_4i_5}\right.\nonumber\\\left.+|H_{j_{i_3}j_{i_2}}(q_{xy})||H_{j_{i_4}j_{i_5}}(q_{xy})|G_{m_{i_3}m_{i_2}}^*(q_z)G_{m_{i_4}m_{i_5}}(q_z)e^{i\frac{\pi}{2}a_2}e^{i\theta(j_{i_2}+j_{i_4}-j_{i_3}-j_{i_5})}\delta K_{i_3i_1i_5i_4}^*\right.\nonumber\\\left.-|H_{j_{i_3}j_{i_1}}(q_{xy})||H_{j_{i_4}j_{i_5}}(q_{xy})|G_{m_{i_3}m_{i_1}}(q_z)G_{m_{i_4}m_{i_5}}^*(q_z)e^{-i\frac{\pi}{2}a_1}e^{-i\theta(j_{i_1}+j_{i_4}-j_{i_3}-j_{i_5})}\delta K_{i_3i_2i_5i_4}\right.\nonumber\\\left.-|H_{j_{i_1}j_{i_3}}(q_{xy})||H_{j_{i_5}j_{i_4}}(q_{xy})|G_{m_{i_1}m_{i_3}}^*(q_z)G_{m_{i_5}m_{i_4}}(q_z)e^{-i\frac{\pi}{2}a_1}e^{-i\theta(j_{i_1}+j_{i_4}-j_{i_3}-j_{i_5})}\delta K_{i_2i_3i_4i_5}^*\right),\nonumber\\
\frac{d}{dt}\delta K_{i_1i_2i_4i_5}=\frac{1}{i\hbar}(E_{i_2}+\hbar\omega_{\text{LO}}-E_{i_1})\delta K_{i_1i_2i_4i_5}-\gamma_{ph}\delta K_{i_1i_2i_4i_5}\nonumber\\+\frac{1}{i\hbar}\left[(n_0+1)n_{i_1i_5}(\delta_{i_4,i_2}-\alpha_Bn_{i_4i_2})-n_0n_{i_4i_2}(\delta_{i_1,i_5}-\alpha_Bn_{i_1i_5})\right],
\label{eq39}
\end{eqnarray}
where the substitution $\alpha_B\delta K_{i_1i_2i_4i_5}\rightarrow \delta K_{i_1i_2i_4i_5}$ is introduced. It follows from Eq.~(\ref{eq37}) that if the condition $j_{i_{1,2}}+j_{i_4}=j_{i_3}+j_{i_5}$ is fulfilled, then $a_{i_{1,2}}=0$. Furthermore, using the identity
\begin{equation}
\int_{\theta=0}^{2\pi}d\theta\;e^{i\frac{\pi}{2}a_{1,2}}e^{\theta(j_{i_{1,2}}+j_{i_4}-j_{i_3}-j_{i_5})}=e^{i\frac{\pi}{2}a_{1,2}}2\pi\delta_{j_{i_{1,2}}+j_{i_4},j_{i_3}+j_{i_5}},
\label{eq40}
\end{equation}
and summing over the phonon wave vector ${\bf q}$, we finally get the expressions given by Eq.~(\ref{eq16}) in Subsec.~\ref{subsec1}.

In order to perform the Markovian approximation, it is assumed that the dominant time dependence is given by the exponential in Eq.~(\ref{eq34}) and therefore the value of electron populations can be taken out of integral.~\cite{kuhn,dm} Also, the fast oscillations of polarizations have to be taken into account,~\cite{kuhn,dm} resulting in
\begin{eqnarray}
\delta s_{k,{\bf q},k-q_y}^{i_1i_2}=-i\pi\sum_{i_4,i_5}g_{\bf q}^*H_{j_{i_5}j_{i_4}}^*(k,k-q_y,q_x)G_{m_{i_5}m_{i_4}}^*(q_z)\delta(E_{i_5}+\hbar\omega_{\text{LO}}-E_{i_4})\nonumber\\\left[(n_0+1)f_{i_1i_5,k}(\delta_{i_4,i_2}-f_{i_4i_2,k-q_y})-n_0f_{i_4i_2,k-q_y}(\delta_{i_1,i_5}-f_{i_1i_5,k})\right]
\label{eq41}
\end{eqnarray}
Following the identical procedure as in the derivation of the quantum kinetic equations, we obtained the equations of motion in the Markov limit given by Eq.~(\ref{eq18}) in Subsec.~\ref{subsec1}. 

\section{}

In this appendix we derive the expression for the lateral overlap integral $H_{j_fj_i}(k_f,k_i,q_x)=\int u_{j_f}^*\left(x-f(k_f)\right)e^{iq_xx}u_{j_f}\left(x-f(k_i)\right)dx$, with the harmonic oscillator wave function of the form 
\begin{equation}
u_j(x)=\sqrt{\frac{\beta}{2^jj!\sqrt{\pi}}}H_j[\beta(x-x_0)]e^{-\frac{\beta^2}{2}(x-x_0)^2},
\label{eq42}
\end{equation}
where $\beta=eB/\hbar$, $x_{0i}=k_i/\beta^2$, $x_{0f}=k_f/\beta^2$, and $k_i=k_f\mp q_y$ (the upper sign holds for phonon absorption and the lower one for emission). Inserting $t=x-x_{0f}$, $t_0=\frac{\mp q_y+iq_x}{2\beta^2}$, and $z=t-t_0$ into Eq.~(\ref{eq42}), the lateral overlap integral may be transformed into
\begin{eqnarray}
H_{j_fj_i}(k_f,k_f\mp q_y,q_x)=\frac{\beta}{\sqrt{\pi}}\frac{1}{\sqrt{2^{j_i+j_f}j_i!j_f!}}e^{iq_xx_{0f}}e^{-\frac{q_{xy}^2}{4\beta^2}}e^{\mp i\frac{q_xq_y}{2\beta^2}}\nonumber\\\int dz\;H_{j_f}(\beta z+\beta t_0)H_{j_i}\left(\beta z+\beta t_0\pm\frac{q_y}{\beta}\right)e^{-\beta^2z^2}.
\label{eq43}
\end{eqnarray}
In order to obtain the final expression for $H_{j_fj_i}(k_f,k_i,q_x)$, we use the following identity
\begin{equation}
\int dx\;e^{-c^2x^2}H_m(a+cx)H_n(b+cx)=\frac{2^n\sqrt{\pi}m!b^{n-m}}{c}L_m^{n-m}(-2ab), n>m.
\label{eq44}
\end{equation}
In the case when $j_i<j_f$, by substituting $c=\beta^2$, $m=j_i$, $n=j_f$, $a=\beta t_0\pm q_y/\beta$, and $b=\beta t_0$, 
Eq.~(\ref{eq43}) takes the form
\begin{eqnarray}
H_{j_fj_i}(k_f,k_f\mp q_y,q_x)=\frac{\beta}{\sqrt{\pi}}\frac{1}{\sqrt{2^{j_i+j_f}j_i!j_f!}}e^{iq_xx_{0f}}e^{-\frac{q_{xy}^2}{4\beta^2}}e^{\mp i\frac{q_xq_y}{2\beta^2}}\frac{2^{j_f}\sqrt{\pi}j_i!(\beta t_0)^{j_f-j_i}}{\beta^2}\nonumber\\L_{j_i}^{j_f-j_i}\left[-2\left(\beta t_0\pm\frac{q_y}{\beta}\right)\beta t_0\right]
\label{eq45}
\end{eqnarray}
Since the following identities
\begin{eqnarray}
-2\left(\beta t_0\pm\frac{q_y}{\beta}\right)\beta t_0=\frac{q_{xy}^2}{2\beta^2},\nonumber\\
\beta t_0=\frac{\mp q_y+iq_x}{2\beta}=\frac{q_{xy}}{2\beta}e^{i\arg(\mp q_y+iq_x)}
\label{eq46}
\end{eqnarray}
hold, the final form of Eq.~(\ref{eq45}) reads
\begin{equation}
H_{j_fj_i}(k_f,k_f\mp q_y,q_x)=|H_{j_fj_i}(q_{xy})|e^{i\frac{q_xk_f}{\beta^2}}e^{\mp i\frac{q_xq_y}{2\beta^2}}e^{i\arg(\mp q_y+iq_x)(j_f-j_i)},
\label{eq47}
\end{equation}
where
\begin{equation}
|H_{j_fj_i}(q_{xy})|=\left(\frac{j_i!}{j_f!}\right)^{\frac{1}{2}}\left(\frac{q_{xy}^2}{2\beta^2}\right)^{\frac{j_f-j_i}{2}}L_{j_i}^{j_f-j_i}\left(\frac{q_{xy}^2}{2\beta^2}\right)e^{-\frac{q_{xy}^2}{4\beta^2}}.
\label{eq48}
\end{equation}
Similarly, if $j_i>j_f$, it can be shown that
\begin{equation}
H_{j_fj_i}(k_f,k_f\mp q_y,q_x)=|H_{j_fj_i}(q_{xy})|e^{i\frac{q_xk_f}{\beta^2}}e^{\mp i\frac{q_xq_y}{2\beta^2}}e^{i\arg(\pm q_y+iq_x)(j_i-j_f)},
\label{eq49}
\end{equation}
with
\begin{equation}
|H_{j_fj_i}(q_{xy})|=\left(\frac{j_f!}{j_i!}\right)^{\frac{1}{2}}\left(\frac{q_{xy}^2}{2\beta^2}\right)^{\frac{j_i-j_f}{2}}L_{j_f}^{j_i-j_f}\left(\frac{q_{xy}^2}{2\beta^2}\right)e^{-\frac{q_{xy}^2}{4\beta^2}}.
\label{eq50}
\end{equation}

\newpage 

\begin{thebibliography}{99}

\bibitem{faist} J. Faist, F. Capasso, D. L. Sivco, C. Sirtori, A. L. Hutchinson, and A. Y. Cho, Science {\bf 264}, 553 (1994).

\bibitem{iottimc} R.~C.~Iotti, and F.~Rossi, Appl. Phys. Lett. {\bf 78}, 2902 (2001).

\bibitem{hansmc} H.~Callebaut, S.~Kumar, B.~S.~Williams, and Q.~Hu, Appl. Phys. Lett. {\bf 84}, 645 (2004).

\bibitem{zoranmc} Z.~Ikoni\'c, R.~W.~Kelsall, and P.~Harrison, Phys. Rev. B {\bf 69}, 235308 (2004). 

\bibitem{zoranre} Z. Ikoni\'c, P. Harrison, and R. W. Kelsall, J. Appl. Phys. {\bf 96}, 6803 (2004). 

\bibitem{dragan} D.~Indjin, P.~Harrison, R.~W.Kelsall, and Z.~Ikoni\'c, J. Appl. Phys. {\bf 91}, 9019 (2002).


\bibitem{iotti} R.~C.~Iotti, and F.~Rossi, Phys. Rev. Lett. {\bf 87}, 146603 (2001).

\bibitem{iotti2} R.~C.~Iotti, and F.~Rossi, Rep. Prog. Phys. {\bf 68}, 2553 (2005).

\bibitem{iotti3} R.~C.~Iotti, E.~Ciancio, and F.~Rossi, Phys. Rev. B {\bf 72}, 125347 (2005).

\bibitem{greenqcl0} A.~Wacker, Phys. Rev. B {\bf 66}, 085326 (2002).

\bibitem{greenqcl} S.~C.~Lee, and A.~Wacker, Phys. Rev. B {\bf 66}, 245314 (2002).

\bibitem{greenqcl2} F.~Banit, S.~C.~Lee, A.~Knorr, and A.~Wacker, Appl. Phys. Lett. {\bf 86}, 041108 (2005).

\bibitem{coherent} S.~C.~Lee, F.~Banit, M.~Woerner, and A.~Wacker, Phys. Rev. B {\bf 73}, 245320 (2006).

\bibitem{inesqcl} I.~Waldm\"{u}ler, W.~W.~Chow, E.~W.~Young, and M.~C.~Wanke, IEEE J. Quantum Electron. {\bf 42}, 292 (2006).


\bibitem{hansdm} H.~Callebaut, and Q.~Hu, J. Appl. Phys. {\bf 98}, 104505 (2005).


\bibitem{ines} I.~Waldm\"{u}ler, J.~F\"{o}rstner, S.~C.~Lee, A.~Knorr, M.~Woerner, K.~Reimann, R.~A.~Kaindl, and T.~Elsaesser, Phys. Rev. B {\bf 69}, 205307 (2004).

\bibitem{ines2} I.~Waldm\"{u}ler, W.~W.~Chow, and A.~Knorr, Phys. Rev. B {\bf 73}, 035433 (2006).

\bibitem{stephan} S.~Butscher, J.~F\"{o}rstner, I.~Waldm\"{u}ler, and A.~Knorr, Phys. Stat. Sol. B {\bf 241}, R49 (2004).

\bibitem{stephan2} S.~Butscher, J.~F\"{o}rstner, I.~Waldm\"{u}ler, and A.~Knorr, Phys. Rev. B {\bf 72}, 045314 (2005).


\bibitem{becker} C.~Becker, C.~Sirtori, O.~Drachenko, V.~Rylkov, D.~Smirnov, and J.~Leontin, Appl. Phys. Lett. {\bf 81}, 2941 (2002).


\bibitem{smirnov1} D. Smirnov, C. Becker, O. Drachenko, V.~V.~Rylkov, H. Page, J. Leotin, and C. Sirtori, Phys. Rev. B {\bf 66}, 121305(R) (2002).

\bibitem{smirnov2} D. Smirnov, O. Drachenko, J. Leotin, H. Page, C. Becker,
 C. Sirtori, V. Apalkov, and T. Chakraborty, Phys. Rev. B {\bf 66}, 125317 (2002).

\bibitem{lobecker} C. Becker, A. Vasanelli, C. Sirtori, and G. Bastard, Phys. Rev. B {\bf 69}, 115328 (2004).

\bibitem{kempa} K.~Kempa, Y.~Zhou, J.~R.~Engelbrecht, P.~Bakshi, H.~I.~Ha, J.~Moser, M.~J.~Naughton, J.~Ulrich, G.~Strasser, E.~Gornik, and K.~Unterrainer,  Phys. Rev. Lett. {\bf 88}, 226803 (2002).

\bibitem{kempa2} K. Kempa, Y. Zhou, J. R. Engelbrecht, and P. Bakshi, Phys. Rev. B {\bf 68}, 085302 (2003).

\bibitem{newmagnqcl} A.~Leuliet, A.~Vasanelli, A.~Wade, G.~Fedorov, D. Smirnov, G.~Bastard, and C.~Sirtori, Phys. Rev. B {\bf 73}, 085311 (2006).

\bibitem{cuba} I.~Savi\'c, P.~Harrison, V.~Milanovi\'c, D.~Indjin, Z.~Ikoni\'c, and V.~D.~Jovanovi\'c, Phys. Stat. Sol. B {\bf 242}, 1812 (2005).

\bibitem{jelena} J.~Radovanovi\'c, V.~Milanovi\'c, Z.~Ikoni\'c, D.~Indjin, and P.~Harrison, J. Appl. Phys. {\bf 97}, 103109 (2005).

\bibitem{prb} I.~Savi\'c, Z.~Ikoni\'c, V.~Milanovi\'c, N.~Vukmirovi\'c ,V.~D.~Jovanovi\'c, D.~Indjin, and P.~Harrison, Phys. Rev. B {\bf 73}, 075321 (2006).

\bibitem{magnqw} K.~El~Sayed, J.~A.~Kenrow, and C.~J.~Stanton, Phys. Rev. B {\bf 57}, 12369 (1998).

\bibitem{magnqw2} M.~W.~Wu, and H.~Haug, Phys. Rev. B {\bf 58}, 13060 (1998).

\bibitem{landau} L.~D.~Landau and E.~M.~Lifshitz, {\it Quantum Mechanics: Nonrelativistic Theory} (Pergamon, London, 1959).

\bibitem{kuhn} T.~Kuhn, in {\it Theory of Transport Properties of Semiconductor Nanostructures}, edited by E.~Sch\"{o}ll (Chapman \& Hall, London, 1998).

\bibitem{dm} T.~Kuhn, and F.~Rossi, Rev. Mod. Phys. {\bf 74}, 895 (2002).


\bibitem{sandra} S.~\v{Z}ivanovi\'c, V.~Milanovi\'c, and Z.~Ikoni\'c, Phys. Rev. B {\bf 52}, 8305 (1995).

\bibitem{haug} H.~Haug, and S.~W.~Koch, {\it Quantum Theory of the Optical and Electronic Properties of Semiconductors} (World Scientific, Singapore, 2004).

\bibitem{jpc} D.~Calecki, J.~F.~Palmier, and A.~Chomette, J. Phys. C {\bf 17}, 5017 (1984).

\bibitem{absspectra} H.~C.~Schneider, W.~W.~Chow, and S.~W.~Koch, Phys. Rev. B {\bf 70}, 235308 (2004).

\bibitem{apl} I.~Savi\'c, Z.~Ikoni\'c, N.~Vukmirovi\'c, D.~Indjin, P.~Harrison, and V.~Milanovi\'c, Appl. Phys. Lett. {\bf 89}, 011109 (2006). 







\end{thebibliography}

\end{document}